\documentclass[prd,twocolumn,reprint,preprintnumbers,nofootinbib,superscriptaddress,longbibliography]{revtex4-1}
\usepackage{eqnarray}
\usepackage{mathrsfs}
\usepackage[scr=rsfs,cal=boondox]{mathalfa}

\usepackage{array,multirow}
\usepackage{float}

\usepackage{amssymb}
\usepackage{enumerate}
\usepackage{amssymb}
\usepackage{amsmath}
\usepackage{tikz}
\usepackage{feynmf}
\usepackage{makecell}
\usepackage{slashed}
\usepackage{natbib}
\usepackage{graphicx}
\usepackage{setspace}
\usepackage{tensor}
\usepackage{physics}
\usepackage{slashed}
\usepackage{pifont}
\usepackage{slashed}
\newcommand{\genericT}{\ensuremath{T}}

\newcommand{\met}{\slashed{E}_\genericT}

\newcommand{\mpt}{\slashed{p}_\genericT}
\newcommand{\mptvec}{\slashed{\vec{p}}_\genericT}

\newcommand{\beq}{\begin{equation}}
\newcommand{\eeq}{\end{equation}}
\newcommand{\bea}{\begin{eqnarray}}
\newcommand{\eea}{\end{eqnarray}}

\newcommand{\pysr}{\textsc{PySR}}
 \newcommand{\amc}{{\sc MadGraph5\textunderscore}a{\sc MC@NLO}}

\begin{document}

\author{Zhongtian Dong}
\email{cdong@ku.edu}
\affiliation{Department of Physics and Astronomy, University of Kansas, Lawrence, KS 66045, USA}

\author{Kyoungchul Kong}
\email{kckong@ku.edu}
\affiliation{Department of Physics and Astronomy, University of Kansas, Lawrence, KS 66045, USA}

\author{Konstantin T. Matchev}
\email{matchev@ufl.edu}
\affiliation{Institute for Fundamental Theory, Physics Department, University of Florida, Gainesville, FL 32611, USA}

\author{Katia Matcheva}
\email{matcheva@ufl.edu}
\affiliation{Institute for Fundamental Theory, Physics Department, University of Florida, Gainesville, FL 32611, USA}

\title{Is the Machine Smarter than the
Theorist: \\ 
Deriving Formulas for Particle Kinematics with Symbolic Regression}

\begin{abstract}
We demonstrate the use of symbolic regression in deriving analytical formulas, which are needed at various stages of a typical experimental analysis in collider phenomenology. As a first application, we consider kinematic variables like the stransverse mass, $M_{T2}$, which are defined algorithmically through an optimization procedure and not in terms of an analytical formula. We then train a symbolic regression and obtain the correct analytical expressions for all known special cases of $M_{T2}$ in the literature. As a second application, we reproduce the correct analytical expression for a next-to-leading order (NLO) kinematic distribution from data, which is simulated with a NLO event generator. Finally, we derive analytical approximations for the NLO kinematic distributions after detector simulation, for which no known analytical formulas currently exist.
\end{abstract}

\date{November 15, 2022}

\maketitle
\tableofcontents

\section{Introduction\label{sec:intro}}

Being able to describe the data collected from the observations of various physical phenomena with simple analytical equations and formulas is the holy grail in theoretical physics --- the physicists who are lucky enough to find such relationships typically get those laws named after them. In the era of big data, this task is becoming increasingly difficult for a human --- the data is just too complex and/or very high-dimensional. Recent advances in computer science and theoretical modelling have allowed us to entertain the idea that the discovery process could perhaps be automated (at least as a matter of principle) and novel laws of phenomenological behavior can be constructed entirely with a machine and without any human intervention \cite{Langley1977, Langley1987,Kokar1986,Langley1989,Zembowicz1992,Todorovski1997,Bongard2007,Schmidt2009,Battaglia2016,Chang2016,Guimera2020}. A less ambitious, but still worthy, task is to simply let the machine re-derive the known classical physics laws from data \cite{Udrescu:2019mnk,Cranmer:2020wew,liu2022ai,https://doi.org/10.48550/arxiv.2206.10540}. 

Spurred by the extensive recent research on symbolic learning in the machine learning community, the above program was recently successfully applied to examples in a wide range of physics areas, e.g. 
in astrophysics \cite{Cranmer2019,Cranmer:2020wew,2021arXiv211102422D}, 
in astronomy for the study of orbital dynamics \cite{Iten2020,Lemos2022}
and exoplanet transmission spectroscopy  \cite{Matchev2022ApJ},
in collider physics \cite{Choi:2010wa,Butter:2021rvz,Dersy:2022bym,Alnuqaydan:2022ncd}, 
in materials science \cite{wang_wagner_rondinelli_2019},
and in behavioral science \cite{Arechiga2021}.
A common ML tool used in such studies is symbolic regression --- an interpretable machine learning algorithm 
which searches the space of functions until it finds an algebraic expression that approximates the dataset well. While most current applications of symbolic regression are limited to low-dimensional data, the approach can be easily extended to higher-dimensional spaces by using a neural network as a proxy, as illustrated in Ref.~\cite{Cranmer:2020wew} with the example of N-body problems.

The basic task in symbolic regression is to learn an analytical expression $f(\mathbf{x})$ given some labelled data $(\mathbf{x},y)$, where $\mathbf{x}$ are input features, typically high-dimensional, and $y$ is the output target label\footnote{In principle, $y$ can also be high-dimensional, however, for simplicity in this paper we shall focus on a single $y$.}. The learned function $f(\mathbf{x})$ can be scrutinized further in three aspects corresponding to fundamental principles of explainable AI \cite{559461}:

\begin{itemize}
\item {\bf Explanation accuracy.} The first question is, how good is the result, i.e., how well does $f(\mathbf{x})$ fit the training data. Typical datasets are imperfect, due to noise, experimental errors, etc., in which case the fitted function will provide only an approximate description of the data. The fit is only expected to get worse as the errors in the data increase \cite{2204.02704}. On the other hand, even if the data is perfect, the fit may be sub-optimal due to factors related to the training of the symbolic regression itself. For example, one may have started with the wrong choice of basis functions, one may have unnecessarily restricted the functional complexity, or the training may simply not converge to the right answer. Our numerical examples considered in this paper shall illustrate many of those situations.
\item {\bf Generalizability (knowledge limits).} A system should only operate under conditions for which it was designed. In the case of symbolic regression, extrapolating into the regions away from the training data in principle could be dangerous and should be handled with care. At the same time, physics laws are universal --- if we find the correct relationship, it should be valid over the full allowed domain of the input variables. As shown below, this principle could be used to narrow down the list of candidate analytical expressions.
\item {\bf Explainability (meaningful).} A system must provide explanations that are understandable to the intended consumers, and furthermore, these explanations must correctly reflect the reason for generating the output and/or the system’s process. A common criticism of deep learning models is that they are black boxes which provide little insight into the fundamental processes that are at work. A symbolic regression is arguably the most intuitive and meaningful approach from the point of view of a theorist --- theorists are used to working with analytical formulas and from experience can often find the physical interpretations of the various terms in an analytical expression.
\end{itemize}

In this paper we consider several applications of symbolic regression to problems in collider physics and specifically particle kinematics. These examples will be presented in order of increasing difficulty, starting from simple cases in which the exact theoretical formula is known. Nevertheless, rederiving those answers with a symbolic regression will serve as an important illustration and validation of our procedure. 

Symbolic regression is a promising machine learning method that searches over a large space of functions until it finds an expression which is both a) relatively simple and b) a good fit to the training data. Because the evolutionary algorithm requires diversity in order to effectively explore the search space, the result of the symbolic regression is a collection of several high-scoring models, which need to be scrutinized by the user to identify an approximation that offers a good trade-off between accuracy and simplicity. At the same time, training a symbolic regression is a computationally expensive process, since the function space to be scanned is in principle infinite. This is why, as a proof-of-concept, in this paper we shall limit ourselves to a few simple examples, which do not require a high-performance cluster, and can be done on a personal laptop. 

To train a symbolic regression, we shall make use of the \pysr~software package \citep{pysr}, which models the data set with a graph neural network before applying symbolic regression to fit different internal parts of the learned model that operate on reduced dimension representations \cite{Cranmer:2020wew}. We shall not attempt any hyperparameter optimization and for the most part will use the default configuration in the \pysr~ version 0.10.1 distribution. 

The paper is organized as follows. In Section~\ref{sec:mt2} we shall use parton-level data (in the narrow-width approximation) to re-derive some known analytical results for the Cambridge $M_{T2}$ variable. In Section~\ref{sec:F} we repeat the same exercise and try to derive the splitting function ${\cal F}(E,\theta)$ for the ISR photon at an $e^+e^-$ collider, which gives us the probability to radiate a photon with a given energy $E$ and a given polar angle $\theta$. We perform two versions of the exercise. First, in Section~\ref{sec:Fdirect} we sample the ${\cal F}$ function directly to create a perfect data sample with no statistical fluctuations. Then, in Section~\ref{sec:F_MC} we use a sample of Monte Carlo (MC) generated events to first obtain a binned estimate of ${\cal F}$ (which is subject to statistical errors) before applying the symbolic regression. In Section~\ref{sec:F_MC_detector} we perform a more realistic analysis by adding detector resolution effects. Section~\ref{sec:conclusions} is reserved for a summary and outlook.

\section{Deriving analytic expressions for algorithmically defined kinematic variables: $M_{T2}$}
\label{sec:mt2}

A standard analysis of particle physics data (such as events from collisions at the Large Hadron Collider (LHC) at CERN) involves the study of distributions of kinematic variables, which are typically defined in terms of the energies and momenta of the particles observed in the detector (for recent reviews of the kinematic variables commonly used in collider phenomenology, see \cite{Han:2005mu,Barr:2010zj,Barr:2011xt,Franceschini:2022vck}). Many of these variables, e.g., invariant mass, missing transverse momentum, etc., are defined in terms of simple analytical expressions and can be readily computed from the collections of particle 4-momenta in the event. However, there also exist another class of kinematic variables, which are defined algorithmically, i.e., through a well-defined optimization procedure which involves the minimization (or maximization) of a relevant kinematic function. In that case, the kinematic variable is a quantity which can be computed only once the algorithm has converged, and typically there is no a priori known analytical expression for it in the general case. Examples of such variables include many traditional event shape variables (thrust, sphericity, etc.) \cite{Banfi:2010xy,Franceschini:2022vck}, some modern substructure variables like N-jettiness \cite{Stewart:2010tn} and N-subjettiness \cite{Thaler:2010tr}, and many others. Another large class of algorithmic variables which have received a lot of attention in the last 15 years, are the so-called constrained mass variables which are computed via constrained minimization of a kinematic function of the particle 4-momenta \cite{Han:2005mu,Barr:2010zj,Barr:2011xt,Franceschini:2022vck}. The minimization is typically performed over the energy and momentum components of invisible particles in the event (neutrinos or dark matter candidates). Examples of constrained mass variables include the Oxbridge variable $M_{T2}$ \cite{Lester:1999tx,Barr:2003rg} and its 4-dimensional generalization $M_2$ \cite{Cho:2014naa,Cho:2015laa,Cho:2014yma}, the variable $M_{2C}$ \cite{Ross:2007rm}, etc. In this paper, for concreteness we shall focus on the well-known $M_{T2}$ variable \cite{Lester:1999tx,Barr:2003rg}, which is algorithmically defined and does not have a known analytical formula in the general case. The advantage of $M_{T2}$ is that there exist formulas for special cases of certain momentum configurations for the visible final state particles. As a warm-up, in this section we shall use these special $M_{T2}$ cases to validate and illustrate the use of symbolic regression for the purpose of deriving new formulas for computing kinematic variables.

\begin{figure}[t]
    \centering
    \includegraphics[width=0.4\textwidth]{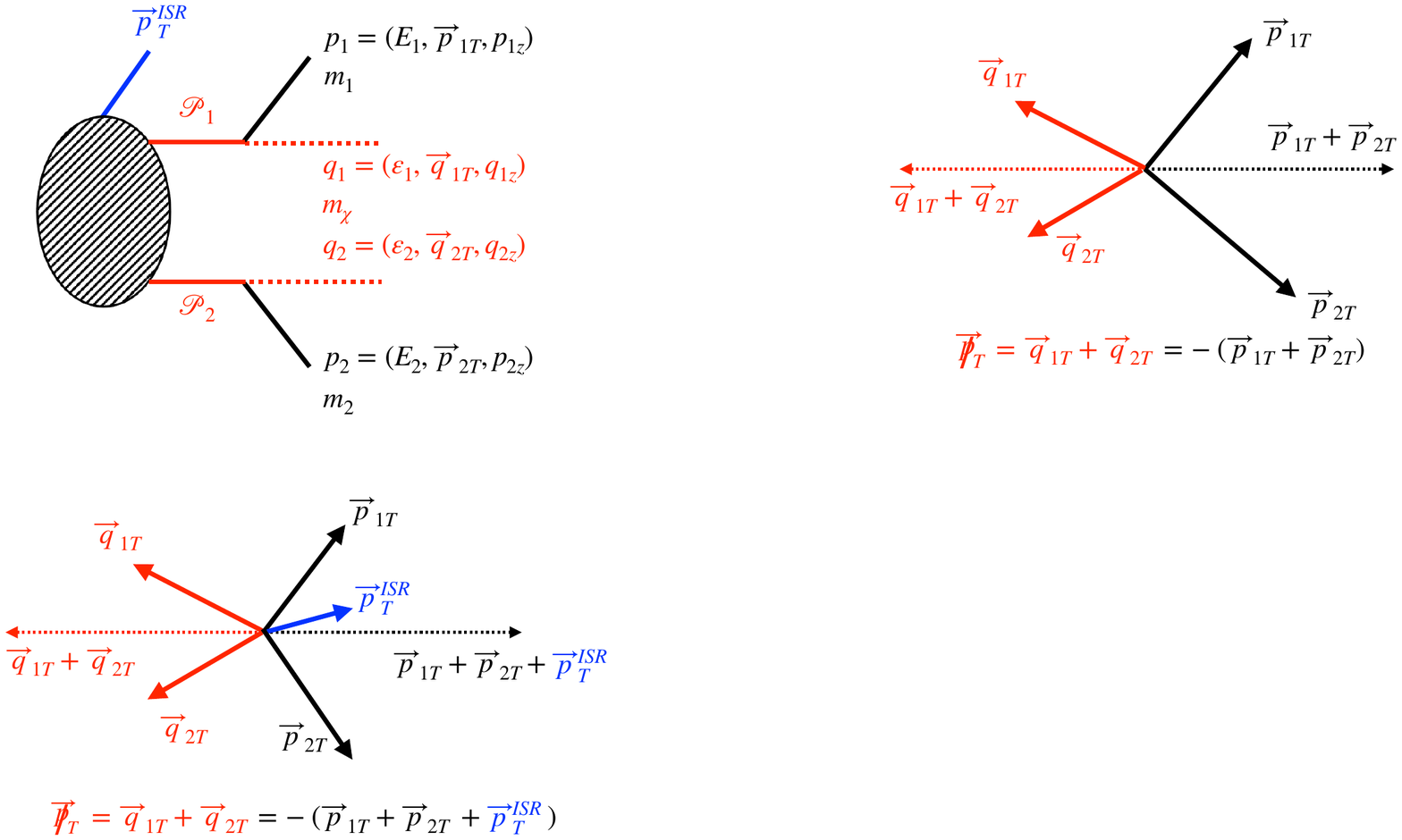}
    \caption{The generic $\met$ event topology applicable to the $M_{T2}$ variable. The parent particles ${\cal P}_1$ and ${\cal P}_2$ are produced in association with some visible upstream transverse momentum $\vec{p}_T^{\, ISR}$. The remaining visible final state particles are divided into two groups (solid black lines), with 4-momenta $p_1$ and $p_2$ and masses $m_1$ and $m_2$, respectively. The two invisible final state particles (red dashed lines) have 4-momenta $q_1$ and $q_2$ and are assumed to have a common mass $m_\chi$. }\label{fig:diagram}
\end{figure}

\begin{figure}[t]
    \centering
    \includegraphics[width=0.45\textwidth]{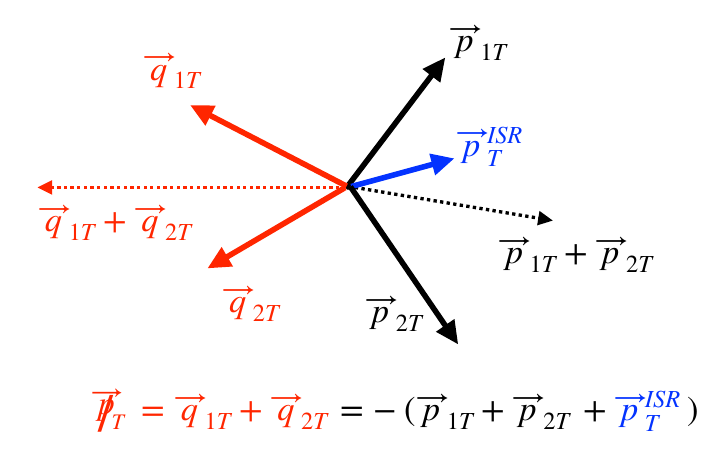}
    \caption{A generic configuration of the transverse momentum vectors $\vec{p}_{iT}$, $\vec{q}_{iT}$, $\mptvec$ and $\vec{p}_{T}^{\,ISR}$ entering the definition (\ref{eq:mt2def}) of $M_{T2}$. 
    }\label{fig:config}
\end{figure}

A well-motivated class of new physics models which generically predict a $\met$ signature, are models with dark-matter candidates. In such models, the lifetime of the dark-matter particle is typically protected by an exact discrete symmetry, which implies that the collider signals will involve not one, but {\it two} decay chains, each terminating in a dark-matter particle invisible in the detector. The simplest $\met$ event topology of this type is illustrated in Figure~\ref{fig:diagram}, where two identical parent particles ${\cal P}_1$ and ${\cal P}_2$ are produced with additional objects, typically from initial state radiation (ISR). Each parent particle ${\cal P}_i$, $(i=1,2)$, decays to a visible particle system with invariant mass $m_i$ and 4-momentum $p_i = (E_i, \vec p_{i T}, p_{i z})$ and an invisible particle $\chi_i$ with 4-momentum $q_i=(\varepsilon_i, \vec q_{i T}, q_{i z})$. The masses of the invisible particles are a priori unknown. Here, we shall assume that the invisible particles $\chi_1$ and $\chi_2$ are identical and have a common mass $m_\chi$. Momentum conservation in the transverse plane implies
\begin{equation}
\vec{q}_{1T} + \vec{q}_{2T} = \mptvec\,,
\label{eq:q1Tq2TMET}
\end{equation}
where the missing transverse momentum vector is given by 
\beq
\mptvec = - (\vec{p}_{1T} + \vec{p}_{2T}) - \vec{p}_{T}^{\,ISR}.
\label{eq:mptdef}
\eeq
The transverse momentum vectors $\vec{p}_{iT}$, $\vec{q}_{iT}$, $\mptvec$ and $\vec{p}_{T}^{\,ISR}$ are illustrated in Figure~\ref{fig:config}. 

The two main ingredients in the $M_{T2}$ calculation are the transverse masses $M_{T{\cal P}_i}$ of the two parent particles ${\cal P}_i$:
\beq
M_{T {\cal P}_i}(\vec q_{iT},m_\chi) = \sqrt{m_i^2 + m_\chi^2  
 + 2 \big ( E_{iT} \varepsilon_{iT}  - \vec p_{iT} \cdot \vec q_{iT} \big )},~~~~
\label{eq:MTPdef} 
\eeq
where the transverse energies are defined as
\beq
E_{i T} = \sqrt{ \vec p_{i T}^{\, 2} + m_i^2  } , \quad
\varepsilon_{i T} = \sqrt{ \vec q_{i T}^{\,2} + m_\chi^2 } \, .
\eeq
The $M_{T2}$ is defined as \cite{Lester:1999tx,Barr:2003rg}
\begin{eqnarray}
M_{T2} (\tilde m) &\equiv& \min_{\vec{q}_{1T},\vec{q}_{2T}}
\hspace*{-0.1cm}
\left\{\max\left[M_{T{\cal P}_1}(\vec{q}_{1T},\tilde m),\,M_{T{\cal P}_2} (\vec{q}_{2T},\tilde m)\right] \right\} ,\nonumber\\
\mptvec  &=& \vec{q}_{1T}+\vec{q}_{2T} \;, \label{eq:mt2def}
\end{eqnarray}
where the {\it a priori} unknown invisible daughter mass $m_\chi$ has been replaced with a test mass parameter $\tilde m$. This construction guarantees that on an event-by-event basis the computed value of $M_{T2}$ does not exceed the mass of the parent ${\cal P}_i$.

In general, the minimization in (\ref{eq:mt2def}) has to be done numerically. However, for certain special cases, analytical solutions have been derived \cite{Barr:2003rg,Lester:2011nj,Lally:2012uj,Lester:2007fq,Cho:2007qv,Cho:2007dh}. In this section, we shall apply symbolic regression to rederive several of those analytical solutions.

\subsection{The case of no upstream momentum}
\label{sec:noISR}

\begin{figure}[t]
    \centering
    \includegraphics[width=0.4\textwidth]{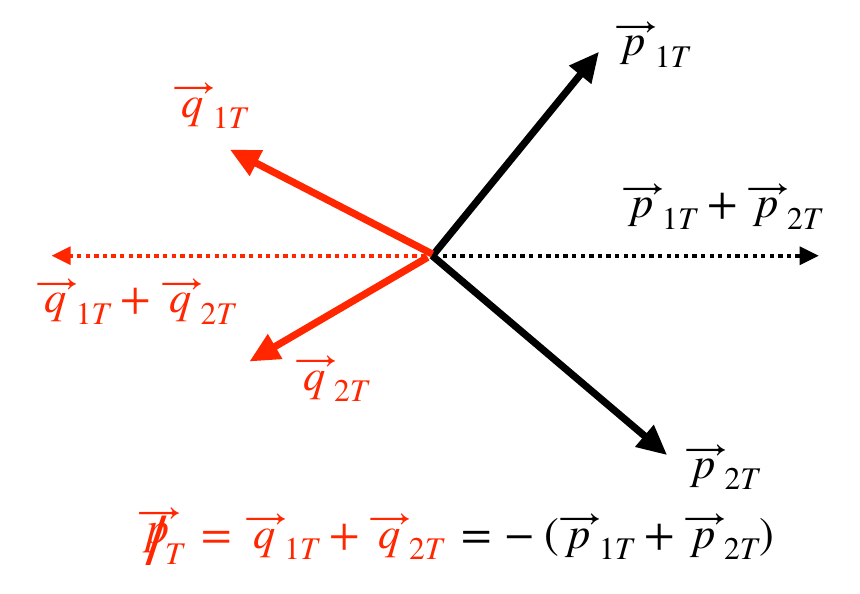}
    \caption{The special momentum configuration with $\vec p_{T}^{\, ISR} = 0$ considered in Section~\ref{sec:noISR}. The missing transverse momentum $\mptvec$ exactly balances the total visible transverse momentum $\vec{p}_{1T}+\vec{p}_{2T}$.
\label{fig:configNoISR}}
\end{figure}

The minimization in eq.~(\ref{eq:mt2def}) may result in one of two distinct possibilities: the transverse masses of the parents are equal, $M_{T {\cal P}_1} = M_{T {\cal P}_2}$, which is known as the balanced solution, or the transverse masses of the parents are unequal, $M_{T {\cal P}_1} \ne M_{T {\cal P}_2}$, known as the unbalanced case. The analytical expression for $M_{T2}$ in the unbalanced case is simply given by eq. (\ref{eq:MTPdef}) \cite{Barr:2003rg}, so the balanced case is the only one we need to worry about. Unfortunately, there is no known analytical formula for the balanced $M_{T2}$ solution for generic momentum configurations like the one in Figure~\ref{fig:config}. However, for the special momentum configuration shown in Figure~\ref{fig:configNoISR}, where $\vec p_T^{ISR}=0$, the analytical formula for the balanced $M_{T2}$ solution is known to be \cite{Cho:2007qv,Lester:2007fq} 
\begin{eqnarray}  
&&M_{T2}^{2}(\tilde{m})
= \tilde{m}^2 + A_T \label{eq:mt2oldB}\\
&&\hspace*{0.5cm}+ 
          \sqrt{ \left ( 1 + \frac{4 \tilde{m}^2}{2A_T-m_1^2-m_2^2} \right ) 
             \left ( A_T^2 - m_1^2 ~m_2^2  \right ) } \,,\nonumber
\end{eqnarray}
where $A_T$ is a convenient shorthand notation introduced in \cite{Cho:2007dh} 
\begin{equation}
A_T = E_{1T} E_{2T} + \vec{p}_{1T}\cdot\vec{p}_{2T} \, .
\label{ATdef}
\end{equation}
In order to avoid always taking an extra square root, from now on for convenience we shall focus on the $M_{T2}$ variable {\em squared}.

In what follows an important attribute of an analytical expression will be the so-called complexity $C$ (defined as the number of leaf nodes in the binary tree representing the analytical expression) \cite{pysr,Cranmer:2020wew}. Clearly, functions of higher complexity in turn will demand more extensive computational resources, including longer computational times. The function (\ref{eq:mt2oldB}) is of complexity 24, which is already a formidable challenge. Given a) our rather modest computational budget, and b) our goal of to demonstrate the method as a proof of principle, here we shall limit ourselves to four simple, yet nontrivial, special cases of (\ref{eq:mt2oldB}) which have lower complexity, namely 
\begin{itemize}
    \item {\em Massless visible and massless invisible final state particles.} Setting $m_1=m_2=0$ and $\tilde m=0$ in (\ref{eq:mt2oldB}), we obtain:
\beq
M_{T2}^2(\tilde m) = 2A_T = 2(E_{1T}\, E_{2 T}+\vec{p}_{1 T} \cdot \vec{p}_{2 T} ) \, .
\label{eq:MT2_special1}
\eeq
    \item {\em Massless visible and massive invisible final state particles.} Substituting $m_1=m_2=0$ and $\tilde m\ne 0$ into (\ref{eq:mt2oldB}), we get:
\beq
M_{T2}^2(\tilde m) = \tilde m^2 + A_T+\sqrt{A_T(A_T+2\tilde m^2) } \, . 
\label{eq:MT2_special2}
\eeq
\item {\em Equally massive visible and massless invisible final state particles.} Alternatively, choosing $m_1=m_2 = m \ne 0$ and $\tilde m= 0$ in (\ref{eq:mt2oldB}), we find:
\beq
M_{T2}^2(\tilde m=0) = A_T + \sqrt{A_T^2- m^4} \, .
\label{eq:MT2_special3}
\eeq    
\item {\em Equally massive visible and massive invisible final state particles.} Finally, choosing $m_1=m_2 = m \ne 0$ and $\tilde m\ne 0$ in (\ref{eq:mt2oldB}), we find:
\beq
M_{T2}^{2}(\tilde{m})
= \tilde{m}^2 + A_T +  \sqrt{ \left ( A_T - m^2 + 2\tilde{m}^2 \right ) 
             \left ( A_T + m^2   \right ) } \, .
\label{eq:MT2_special4}
\eeq    
\end{itemize}
We shall now try to reproduce\footnote{ Note that all of these results would have been completely novel prior to 2007, i.e., only 15 years ago.} each of those expressions (\ref{eq:MT2_special1}-\ref{eq:MT2_special4}) with the symbolic regression algorithm implemented in \pysr~\cite{pysr,Cranmer:2020wew}. For this purpose, we shall generate a large sample of events, compute the target variable $M_{T2}^2$ numerically from the defining formula (\ref{eq:mt2def}), using the python code {\sc mt2 1.2.0} \cite{Lester:2014yga}, and then ask the symbolic regression to ``discover" the analytical results (\ref{eq:MT2_special1}-\ref{eq:MT2_special4}).

\begin{table*}[t]
\centering
\renewcommand\arraystretch{2.0}
\begin{tabular}{||c|c|c|c|c||}
\hline
Case &Complexity & Fitted function & MSE & Score \\
\hline
\hline
\parbox[t]{2mm}{\multirow{3}{*}{\rotatebox[origin=c]{90}{$\mu =0$, $\tilde \mu=0$}}}
& 1 & $\vec{p}_{T1} \cdot \vec{p}_{T2}$ & $7\times10^7$& $0 $\\
& 3 &$|\vec{p}_{T1}+ \vec{p}_{T2} |^2$ & $2.2\times10^6 $&$1.74$ \\
& 5 & $2(\vec{p}_{T1} \cdot \vec{p}_{T2}+E_{T1}\, E_{T2} )$& $0$ & $\infty$\\
\hline\hline
\parbox[t]{2mm}{\multirow{4}{*}{\rotatebox[origin=c]{90}{$\mu =0$, $\tilde \mu \ne 0$}}}
& 9 & $2A_T+1.8(\tilde\mu-3.91) $ & $7.345\times10^3$ & $6.73\times10^{-3}$ \\
& 11 & $2A_T+\tilde\mu/0.556-9.51 $ & $7.316\times10^3 $ & $1.985\times10^{-3}$ \\
& 13 & $2A_T+\tilde\mu+0.10\,\tilde\mu\, A_T^{1/4} $ & $6.377\times10^3 $ & $6.870\times10^{-2}$  \\
& 14 & $\tilde\mu+A_T+\sqrt{A_T^2+2\tilde\mu A_T} $ & $0$ & $\infty$ \\
\hline\hline
\parbox[t]{2mm}{\multirow{4}{*}{\rotatebox[origin=c]{90}{$\mu \ne 0$, $\tilde \mu= 0$}}}
& 5 & $2.02(A_T- 133.35) $ & $6.67\times 10^4$ & $0.23$ \\
& 7 & $2.03A_T-0.44\mu $ & $3.39\times 10^4$ & $ 0.34$\\
& 9 & $2A_T -\mu^2/A_T$ & $ 2.06\times 10^4$ & $0.25$\\
& 10 & $A_T + \sqrt{(A_T-\mu)(A_T+\mu)}$ & $0$ & $\infty $\\
\hline\hline
\parbox[t]{2mm}{\multirow{4}{*}{\rotatebox[origin=c]{90}{$\mu \ne 0$, $\tilde \mu\ne 0$}}}
& 12 & $ A_T +  \sqrt{2}\sqrt{ \left ( A_T - \mu+\tilde{\mu} \right ) 
              A_T }$ & $3.90\times10^4$ & $0.28$ \\
& 13 & $ A_T +  \sqrt{2}\sqrt{ \left ( A_T - 0.99\mu \right ) 
              A_T } $  &  $3.34\times10^4$ & $ 0.16$\\
& 14 & $ A_T +  \sqrt{ \left ( A_T - \mu + 2.74\tilde{\mu} \right ) 
             \left ( A_T + \mu   \right ) } $  & $3.488\times10^3$ & $2.26$\\
& 16 & $\tilde{\mu} + A_T +  \sqrt{ \left ( A_T - \mu + 2\tilde{\mu} \right ) 
             \left ( A_T + \mu   \right ) } $ & $1.12\times10^{-6}$ & $10.93$\\
\hline\hline
\end{tabular}
\caption{Results from the $M_{T2}$ exercise with no ISR considered in Section~\ref{sec:noISR}.
In each case, we show the best fitted functions at several representative values of the complexity. The correct answers are given by eqns.~(\ref{eq:MT2_special1}-\ref{eq:MT2_special4}), with
the substitutions $\tilde m^2 \to \tilde\mu$ and $m^2\to \mu$.
\label{tab:noISR}
}
\end{table*}

In the case of no upstream momentum ($\vec p_T^{\,ISR}=0$) considered in this subsection, there are 7 input degrees of freedom, which naively can be taken to be the two transverse momentum components of each visible particle, $\vec{p}_{1T}$ and $\vec{p}_{2T}$, their masses $m_1$ and $m_2$, and the invisible test mass $\tilde m$. In principle, one can use this set of primitive variables as inputs to the symbolic regression, but the disadvantage is that the machine will need to learn the physics principles from scratch. In order to improve and speed up the performance of the symbolic regression, it is crucial to use an optimized set of input variables which reflects the underlying physics principles of the problem. One possibility is to use dimensional analysis and feed only groups of variables which have the proper physics dimensions \cite{Matchev2022ApJ}. In our case here, since we are looking for a formula for a mass-squared quantity, $M_{T2}^2$, it makes sense if all of our input variables have mass-dimension 2, otherwise the complexity of the function will increase, making it more difficult for the symbolic regression to find it. Furthermore, we know that the answer must be rotationally invariant, back-to-back boost invariant \cite{Cho:2007qv,Cho:2007dh}, and symmetric with respect to permutations among the visible particles $(1\leftrightarrow 2)$. These considerations restrict the relevant set of variables to fewer degrees of freedom, which further improves the performance of the symbolic regression. For example, in the case of the function (\ref{eq:MT2_special1}) we shall consider as inputs the set
$\{E_{1T}E_{2T}, \vec{p}_{1T}\cdot \vec{p}_{2T},|\vec{p}_{1T}+ \vec{p}_{2T}|\}$,
in terms of which the answer (\ref{eq:MT2_special1}) is only of complexity 5. Similarly, in the case of the function (\ref{eq:MT2_special2}), we shall input the values of $A_T$ and $\tilde\mu \equiv\tilde m^2$, which results in complexity 12. Then for the function (\ref{eq:MT2_special3}) we shall use the values of $A_T$ and $\mu \equiv m^2$ as inputs, and the corresponding complexity is 8. Finally, for the function (\ref{eq:MT2_special4}) we shall feed in $\{A_T, \mu, \tilde\mu\}$ and the complexity is 15. 

\begin{table*}[t]
\centering
\renewcommand\arraystretch{2.0}
\begin{tabular}{||c|c|c|c|c||}
\hline
Case & Complexity & Fitted function & MSE & Score \\
\hline
\hline
\parbox[t]{2mm}{\multirow{4}{*}{\rotatebox[origin=c]{90}{$\mu =0$, $\tilde \mu\ne0$}}}
& 8 & $\sqrt{\tilde\mu(\vec{p}_{T1} \cdot \vec{p}_{T2}+E_{T1}\, E_{T2})/0.468}$ & $26.9$ & $ 2.946$ \\
& 10 & $\tilde\mu+\sqrt{\tilde\mu\,(\vec{p}_{T1} \cdot \vec{p}_{T2}+E_{T1}\, E_{T2})/0.5}$ & $2.91\times10^{-5} $ & $ 6.87$ \\
& 12 & $\tilde\mu+\sqrt{2\tilde\mu(\vec{p}_{T1} \cdot \vec{p}_{T2}+E_{T1}\, E_{T2})} +0.005$ & $ 1.25\times10^{-5}$ & $4.24\times10^{-1}$ \\
& 13 & $\tilde\mu+(\sqrt{\tilde\mu} )\sqrt{\vec{p}_{T1} \cdot \vec{p}_{T2}+E_{T1}\, E_{T2}}\sqrt{2}$ & $0$ & $\infty$ 
\\
\hline\hline
\parbox[t]{2mm}{\multirow{4}{*}{\rotatebox[origin=c]{90}{$\mu \ne 0$, $\tilde \mu\ne0$}}}
& 6 & $\sqrt{\tilde\mu\, A_T/0.22}$ & $ 5.33\times10^5$& $ 9.168\times10^{-1}$\\
& 8 &$(\mu+\sqrt{\tilde\mu\, A_T/0.296})$ & $ 1.64\times10^5$ & $5.89\times10^{-1} $ \\
& 10 & $1.29(\tilde\mu+\mu+\sqrt{\tilde\mu  A_T})$ & $ 7.08\times10^{3}$ & $ 1.57$ \\
& 12 & $\tilde\mu+\mu+\sqrt{2\tilde\mu(\mu+A_T)}$ & $0$ & $\infty$ 
\\
\hline\hline
\end{tabular}
\caption{Results from the $M_{T2}$ exercise considered in Section~\ref{sec:noMET} for the momentum configuration with $\mpt=0$ displayed in Figure~\ref{fig:configNoMET}.
\label{tab:noMET}
}
\end{table*}

In order to train the symbolic regression, we need to create suitable training data. For the exercise in this subsection, we sample the lepton momenta $\vec{p}_{1T}$ and $\vec{p}_{2T}$, which also fixes the missing transverse momentum vector as $\mptvec = - (\vec{p}_{1T}+\vec{p}_{2T})$
(see Figure~\ref{fig:configNoISR}). From those momenta we compute the input features (of mass dimension 2) to the symbolic regression as explained above. The target variable $M_{T2}^2$ is then calculated numerically with the {\sc mt2} code \cite{Lester:2014yga}. This exercise is performed four different times, depending on the choice for the mass parameters $\mu$ and $\tilde \mu$ being zero or non-zero, leading to the four different cases in eqns.~(\ref{eq:MT2_special1}-\ref{eq:MT2_special4}).

In each of these four cases, we train the \pysr~symbolic regression algorithm on 10,000 events. We mostly use the default hyperparameter configuration in the \pysr~distribution. Due to the relatively high complexity of our functions, we increased the number of iterations to 10. We allow for the simple arithmetic operators addition ($+$), subtraction ($-$), multiplication ($*$), division ($/$), and square root ($\sqrt{\cdot}$). The loss function is the mean squared error (MSE).

The output from a typical \pysr~run is a set of functions of increasing complexity $C$, together with their MSE and score. 
The score is calculated by the fractional drop in the MSE over the increase in the complexity from the next best model \cite{Cranmer:2020wew}
\begin{eqnarray}
{\rm Score} &=& - \frac{\Delta \log \left( MSE \right )}{\Delta c} \, .
\end{eqnarray}

The results from the four exercises in this subsection are displayed in Table~\ref{tab:noISR}. In each case, the symbolic regression was able to eventually reproduce the correct functional dependence, once the required complexity was reached. We note that Eq.~(\ref{eq:MT2_special4}) turned out to be more challenging than the others, so for that case we increased the population{\underline{\hspace*{0.2cm}}}size parameter to 50 and used 40,000 training samples with batchsize 5,000.

Note that sometimes we obtain an equivalent expression of slightly higher complexity. For example, in the case of eq.~(\ref{eq:MT2_special2}), the answer has expanded the parentheses under the square root, which leads to an equivalent expression, but formally increases the complexity of the function to 14. Also note that since in this exercise the data is sampled from the exact function (no noise or errors), the MSE for the right answer is zero (or very close to it) and the score is infinite (or very large). The successful replication of the known special cases (\ref{eq:MT2_special1}-\ref{eq:MT2_special4}) validates our use of symbolic regression as implemented in \pysr~and motivates us to consider more realistic examples in the following sections.

\subsection{The case of no missing transverse momentum}
\label{sec:noMET}

\begin{figure}[t]
    \centering
    \includegraphics[width=0.45\textwidth]{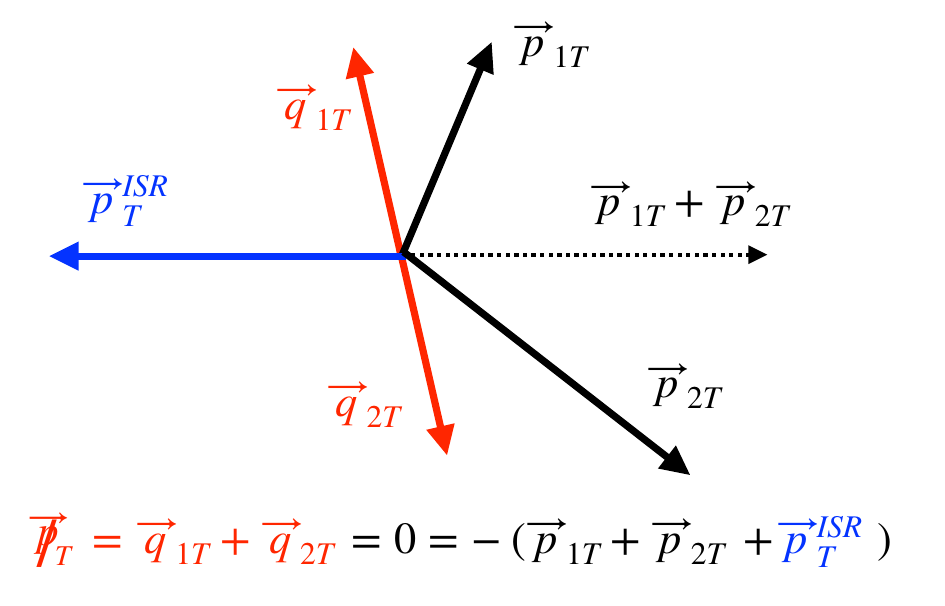}
    \caption{The special momentum configuration with $\mptvec = 0$ considered in Section~\ref{sec:noMET}. The invisible particles have equal and opposite momenta in the transverse plane. As a result, $\vec{p}_T^{\, ISR}$ exactly balances the total visible transverse momentum $\vec{p}_{1T}+\vec{p}_{2T}$. 
    \label{fig:configNoMET}}
\end{figure}

\begin{table*}[t]
\centering
\renewcommand\arraystretch{2}
\begin{tabular}{||c|c|c|c|c||}
\hline
Case & Complexity & Fitted function & MSE & Score \\
\hline
\hline
\parbox[t]{2mm}{\multirow{3}{*}{\rotatebox[origin=c]{90}{$\mu =0$, $\tilde \mu = 0$}}}
& 3 & $|\mptvec||\vec{p}_{T12}|$ & $3.13\times10^{5}$ & $1.59$ \\
& 5 & $|\mptvec|(|\vec{p}_{T12}|-4.04)$ & $2.11\times10^{5}$ & $0.20$ \\
& 7 & $ 2A_T|\mptvec|/|\vec{p}_{T12}|$ & $8.65\times10^{-8}$ & $14.26$ \\
\hline\hline
\parbox[t]{2mm}{\multirow{3}{*}{\rotatebox[origin=c]{90}{$\mu =0$, $\tilde \mu\ne0$}}}
& 14 & $\tilde\mu-QA_T+\sqrt{A_T(\tilde\mu+A_T)/0.50}$ & $52.83$ & $1.08$ \\
& 16 & $\tilde\mu-QA_T/0.897+\sqrt{A_T(2\tilde\mu+A_T)}$ & $44.51$ & $8.57\times10^{-2}$ \\
& 18 & $\tilde\mu-QA_T+\sqrt{A_T(2\tilde\mu+Q^2A_T)} $ & $3.86\times10^{-5}$ & $6.98$ \\
\hline\hline
\end{tabular}
\caption{
Results from the $M_{T2}$ exercise with the collinear momentum configuration in Section~\ref{sec:collinear}.
\label{tab:collinear}
}
\end{table*}

Recently, Ref.~\cite{Lester:2011nj} pointed out a new special case which also allows an analytical formula for $M_{T2}$. Its momentum configuration is shown in Figure~\ref{fig:configNoMET}, where the two invisible momenta are equal and opposite, and as a result $\mptvec=0$. This cancellation of the invisible momenta is purely accidental, which is why this case is mostly of academic interest --- there will be very few events (if any) of this type in the data. Nevertheless, for completeness we shall explore this situation as well.

For simplicity, we shall focus on the case when the masses of the visible final state particles are the same, i.e., $m_1=m_2\equiv m$. The formula for $M_{T2}$ is given by \cite{Lester:2011nj}
\begin{subequations}
\bea
M_{T2}^2(\tilde \mu) &=& \tilde\mu+\mu+\sqrt{2\tilde\mu(\mu+E_{1T} E_{2T}+\vec{p}_{1T} \cdot \vec{p}_{2T})}~~~~~~ \label{eq:MT2MET0_pE} \\
&=& \tilde\mu+\mu+\sqrt{2\tilde\mu(\mu+A_T)}, \label{eq:MT2MET0_AT}
\eea
\label{eq:MT2MET0}
\end{subequations}
where to facilitate later comparisons to the \pysr~output, we have used the mass squared parameters $\tilde\mu=\tilde m^2$ and $\mu=m^2$.

Once again, we may consider several special cases, depending on the masses of the visible and invisible particles. The case of massless invisible particles ($\tilde \mu=0$) leads to a trivial function $M^2_{T2}=\mu$ and will not be considered further. On the other hand, the case of massless visible particles ($\mu=0$ in (\ref{eq:MT2MET0_pE})) gives a non-trivial function
\beq
M_{T2}^2(\tilde \mu) = \tilde\mu+\sqrt{2\tilde\mu(E_{1T} E_{2T}+\vec{p}_{1T} \cdot \vec{p}_{2T})}\, .
\label{eq:MT2MET0_speical1}
\eeq
Keeping in mind that the answer must be symmetric with respect to interchanging $1\leftrightarrow 2$, we can use the set of mass-dimension 2 variables $\{E_{1T}E_{2T}, \vec{p}_{1T}\cdot \vec{p}_{2T}, \tilde\mu\}$, in terms of which the function (\ref{eq:MT2MET0_speical1}) is of complexity 10. 

Proceeding as in Section~\ref{sec:noISR}, we train a symbolic regression  with the default parameter configuration in {\sc PySR} on 10,000 events in the $\mptvec=0$ configuration of Figure~\ref{fig:configNoMET}. We repeat the exercise twice --- once for massless visible particles ($\mu=0$) and then again for massive visible particles ($\mu\ne 0$). The value of $M_{T2}$ is always computed with massive invisible particles ($\tilde\mu\ne 0$). The results are displayed in Table~\ref{tab:noMET}. In the case $\mu=0$, the exact formula (\ref{eq:MT2MET0_speical1}) is reproduced, albeit in a mathematically equivalent form of slightly higher complexity. In the massive case ($\tilde\mu\ne 0$), our set of input variables was taken to be $\{\mu, \tilde\mu, A_T\}$, and the result was again successful, reproducing the correct function (\ref{eq:MT2MET0_AT}) at complexity level 12.

\subsection{The collinear momentum configuration}
\label{sec:collinear}

\begin{figure}[t]
    \centering
    \includegraphics[width=0.45\textwidth]{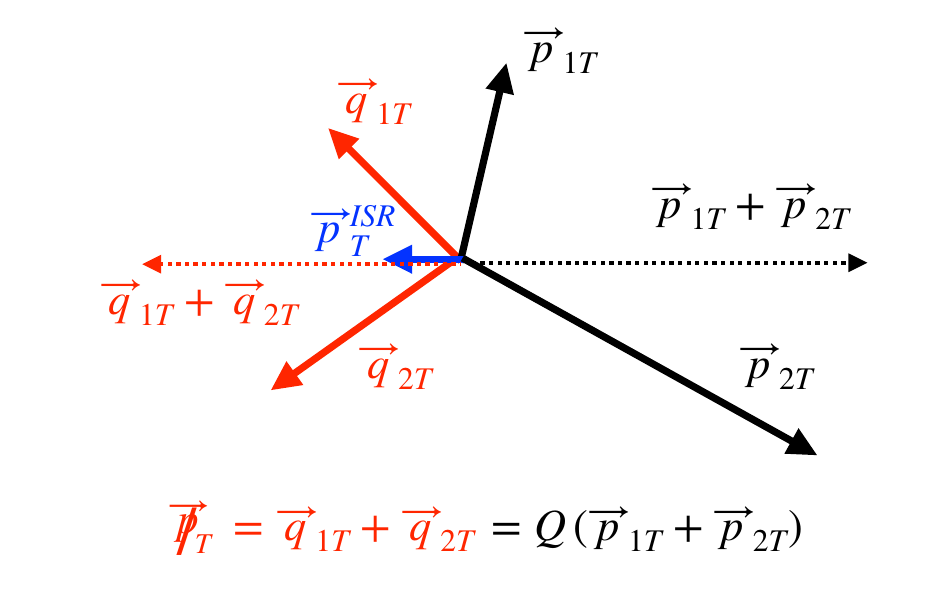}
    \caption{The special balanced momentum configuration $\mptvec = Q ( \vec p_{1T} + \vec p_{2T})$ considered in Section~\ref{sec:collinear}. In general, the proportionality factor $Q$ can be positive or negative, while $Q=0$ reduces to the case considered in Section~\ref{sec:noISR}. Note that $\vec{p}_{T}^{\,ISR}$ is also necessarily collinear with $\mptvec$ and the total visible transverse momentum $\vec{p}_{1T}+\vec{p}_{2T}$.
    \label{fig:configCollinear}}
\end{figure}

A second special case discussed in Ref.~\cite{Lester:2011nj} is that of the collinear momentum configuration in Figure~\ref{fig:configCollinear}, where the three transverse vectors $\mptvec$, $\vec{p}_T^{ISR}$ and $\vec{p}_{1T}+\vec{p}_{2T}$ all lie along the same line in the transverse plane. Ref.~\cite{Lester:2011nj} parametrized this case through a proportionality factor $Q$ defined by
\beq
\mptvec = Q\, (\vec{p}_{1T}+\vec{p}_{2T}) \equiv Q\, \vec{p}_{T12},
\label{eq:Qfactordef}
\eeq
where we have introduced a shorthand notation $\vec{p}_{T12}$ for the total visible transverse momentum $\vec{p}_{1T}+\vec{p}_{2T}$. Note that $Q$ is unbounded and can take both positive and negative values, i.e., $-\infty < Q < \infty$. For definiteness, in Figure~\ref{fig:configCollinear} we show the case of $Q<0$.

Like before, we only consider the case of $\mu=0$ in which case the formula is
\beq
M_{T2}^2(\tilde \mu) = \tilde\mu-QA_T+\sqrt{A_T(2\tilde\mu+Q^2A_T)} .
\eeq
For completeness we also consider separately the special case $\tilde\mu=0$ when the formula simplifies to
\bea
M_{T2}^2(\tilde \mu) &=& - QA_T + |QA_T| \nonumber \\
&=& \left\{ 
\begin{array}{ll}
0,     &  {\rm for~} Q>0,\\
2 A_T |Q| = 2 A_T \frac{|\mptvec|}{|\vec{p}_{T12}|},     & {\rm for~} Q<0.
\end{array}
\right.
\label{ATQ}
\eea
For simplicity we shall only test the non-trivial case given by the second line in (\ref{ATQ}).

Training \pysr~as before, we find the results shown in Table~\ref{tab:collinear}. In the case of $\tilde\mu= 0$, we choose the variables $A_T$, $|\mptvec|$ and $|\vec{p}_{T12}|$ as our input features, while in the case of $\tilde\mu\ne 0$, our input features were $A_T$, $Q$, and $\tilde \mu$. We see that the correct answers are reproduced at complexities 7 and 18, respectively.

\section{Deriving analytic expressions for NLO kinematic distributions}
\label{sec:F}

As our second example, we shall apply symbolic regression to learn the shapes of kinematic distributions at next-to-leading order (NLO). For simplicity, we shall consider the simplest possible process at leading order (LO), namely, the pair-production $e^+e^-\to \chi\chi$ of two invisible particles at a lepton collider with CM energy $\sqrt{s}$. Here the $\chi$ particles can be neutrinos or stable BSM dark matter candidates that escape undetected. In order to observe such events, we have to tag with a photon from initial state radiation (ISR), i.e., consider the NLO process $e^+e^-\to\chi\chi+\gamma$ \cite{Birkedal:2004xn}. 

In general, there is no model-independent exact theoretical prediction for the resulting kinematic distribution of the ISR photon (for model-dependent studies, see \cite{Gopalakrishna:2001iv,Oller:2004br,Mawatari:2014cja,Kalinowski:2022xjw}). However, if the emitted photon is either {\it soft} or {\it collinear} with the incoming electron or positron, soft/collinear factorization theorems provide an approximate model-independent relation between the LO and NLO differential cross-sections:
\beq
\frac{d\sigma(e^+e^-\to \chi\chi+\gamma)}{dx\, d\cos\theta} \approx
{\cal F}(x, \sin\theta)\,\hat{\sigma}(e^+e^-\to\chi\chi),
\label{collinear}
\eeq
where $\theta$ is the angle between the photon direction and the direction of the incoming electron beam, and the dimensionless quantity 
\beq
x=\frac{2E_\gamma}{\sqrt{s}}
\label{xdef}
\eeq
is a measure of the photon energy $E_\gamma$, normalized by the beam energy $\sqrt{s}/2$. Further, $\hat{\sigma}$ is the LO $\chi$ pair-production cross section evaluated at the reduced center of mass energy, $\hat{s}=(1-x)s$. Finally, ${\cal F}$ denotes the splitting function
\beq
{\cal F}(x, \sin\theta) = \frac{\alpha}{\pi}\frac{1+(1-x)^2}{x} 
\frac{1}{\sin^2\theta}\,, \label{llog}
\eeq
which upon integration over $\theta$, reproduces the familiar Weizsacker-Williams distribution function. The factor ${\cal F}$ is universal: it does not depend on the nature of the (electrically neutral) particles produced in association with the photon. 

\begin{table}[t]
\scalebox{0.97}{
\centering
\renewcommand\arraystretch{1.7}
\begin{tabular}{||c|c|c|c||}
\hline
Complexity & Fitted function & MSE & Score \\
\hline
\hline
5 & $ (3.73)/\sin^2\theta$ & $5.56\times10^{3}$ & $0.14$ \\
7 & $ 1.60/(x\sin^2\theta)$ &$2.08\times10^{2}$ & $1.64$ \\
9 &$ (-1.15+\frac{1.89}{x})/\sin^2\theta$  & $8.63$ & $1.59$ \\
11 &$ (x-2+\frac{2}{x})/\sin^2\theta$  & $5.71\times10^{-11}$ & $12.87$\\
\hline\hline
\end{tabular}
}
\caption{Results from the warm-up symbolic regression exercise considered in Section~\ref{sec:Fdirect}. 
\label{tab:Fdirect_2}
}
\end{table}

\begin{table*}[t]
\centering
\renewcommand\arraystretch{1.7}
\begin{tabular}{||c|c|c|c||}
\hline
Complexity & Fitted function & MSE & Score \\
\hline
\hline
9 & $ (-0.039+0.063/x)/\sin^2\theta$ & $8.32\times10^{-3}$ & $1.59$ \\
11 & $ [-0.030+0.063/(x-0.012)]/\sin^2\theta$ &$2.72\times10^{-3}$ & $0.558$ \\
13 &$ (-0.068+0.068/x+0.034x)/\sin^2\theta$  & $1.53\times10^{-4}$ & $1.44$ \\
15 &$ [(-0.067+0.067/x+0.034x)/\sin\theta-0.001]/\sin\theta$  & $1.51\times10^{-4}$ & $6.80\times10^{-3}$\\
\hline\hline
\end{tabular}
\caption{Results for a few representative complexities from the symbolic regression exercise performed in Section~\ref{parton-splitting}. 
\label{tab:Fdirect}
}
\end{table*}

Note that the normalization of (\ref{collinear}) depends on the fine structure coupling constant $\alpha$ appearing in (\ref{llog}). Our main goal in this section will be to apply symbolic regression and learn {\em the shape} of the splitting function (\ref{llog}) from a sample of MC events generated either according to the soft/collinear approximation (\ref{collinear}) (see Sections ~\ref{parton-splitting} and \ref{detector-splitting}) or using the full matrix element in a specific model (see Sections ~\ref{parton-madgraph} and \ref{detector-madgraph}). In Section~\ref{sec:F_MC} (Section~\ref{sec:F_MC_detector}) the exercise will be performed without (with) detector effects, i.e., smearing the photon energy according to the calorimeter resolution.

\subsection{Warm-up toy exercise: learning the splitting function directly}
\label{sec:Fdirect}

\begin{figure}[t]
\centering
\includegraphics[width=0.45\textwidth]{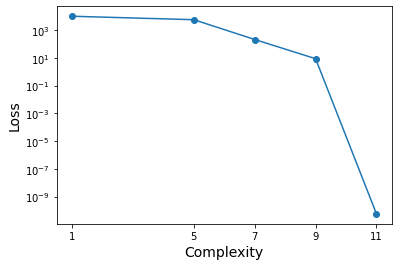}
    \caption{The MSE loss as a function of complexity for the warm-up symbolic regression exercise considered in Section~\ref{sec:Fdirect}. 
    }
    \label{fig:lossIIIA}
\end{figure}

First we begin with a toy exercise where we create the training data by sampling the function ${\cal F}$ directly, i.e., for a given choice of $x$ and $\theta$, we compute the target variable $y$ directly from eq.~(\ref{llog}). In other words, our training dataset will be the set
\beq
\left(x, \sin\theta, \frac{{\cal F}(x, \sin\theta)}{\alpha/\pi}\right),
\label{trainingdata}
\eeq
where for simplicity we have factored out the constant $\alpha/\pi$. One can view this exercise as corresponding to the case of infinite MC statistics in the absence of any detector effects.

We generate training data (\ref{trainingdata}) by sampling $x\in [0.1, 1]$ and $\sin\theta\in[0.1, 1]$ on a $100\times100$ grid. Using the default parameter options in \pysr, we obtain the results shown in Table~\ref{tab:Fdirect_2} for the target function in this case, $\frac{\pi}{\alpha}{\cal F}(x, \sin\theta)$. We see that the correct analytical expression, $(1+(1-x)^2)/(x\sin^2\theta)$, is recovered at complexity 11, as indicated by the sharp drop of the MSE loss (note also the drastic improvement in the score at complexity 11). This is pictorially illustrated in Figure \ref{fig:lossIIIA}, which shows the evolution of the MSE loss as a function of complexity.

\subsection{Learning from gen-level MC data}
\label{sec:F_MC}

Having validated our symbolic regression procedure with the toy example of the previous subsection, we shall now  modify this exercise, making it more realistic in several ways:
\begin{itemize}
    \item Instead of considering infinite statistics, we shall now limit ourselves to a finite event sample, thereby introducing statistical errors in the target values of the function which are used for the training of the symbolic regression. 
    \item In the toy example of Section~\ref{sec:Fdirect}, we generated the training data by simply looking up the value of the target function at a given $x$ and $\sin\theta$ from the correct formula. In reality this will be impossible, and the target values in the training data will have to be determined from experimental or MC simulated data via some sort of density estimation, e.g., through bin counts. Therefore, from now on we shall always rely on MC simulated data to obtain the values for the (unit-normalized) target function 
    from event counts in suitably chosen bins. This approach is a better representation of what would be done in an actual experiment.
    \item While in this subsection we shall restrict ourselves to gen-level data, in the next subsection \ref{sec:F_MC_detector} we shall account for the finite detector resolution by smearing the photon energy.
\end{itemize}

\subsubsection{Data generated using a splitting function}
\label{parton-splitting} 

For this version of the symbolic regression exercise, we first generate MC data according to the approximate model-independent differential cross-section (\ref{collinear}). We avoid the soft/collinear singularity at $x=0$ and $\sin\theta=0$ by focusing on the previously considered region $x\in [0.1, 1]$ and $\sin\theta\in[0.1, 1]$ binned on a $100\times100$ grid. We then sample 100 million events and populate the bins, whose final event counts then serve as the values of the target function (after unit-normalization) to be used in the training. The input features are again $x$ and $\sin\theta$ and we use the default parameter setup in \pysr.

\begin{figure}[t]
\centering
\includegraphics[width=0.47\textwidth]{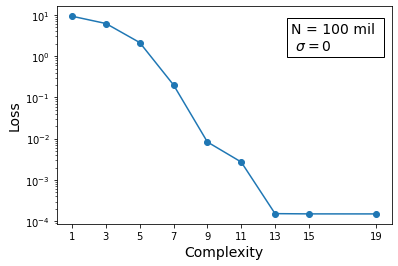}
    \caption{The same as Figure~\ref{fig:lossIIIA}, but for the symbolic regression exercise performed in Section~\ref{parton-splitting}. 
    }\label{fig:loss}
\end{figure}

\begin{figure}[t]
    \centering
    \includegraphics[width=0.23\textwidth]{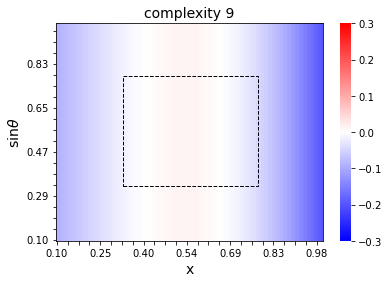}
    \includegraphics[width=0.23\textwidth]{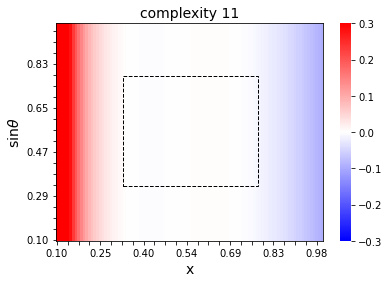} \\
    \includegraphics[width=0.23\textwidth]{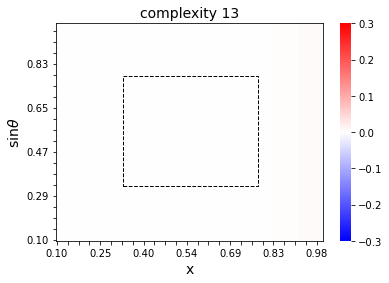}
    \includegraphics[width=0.23\textwidth]{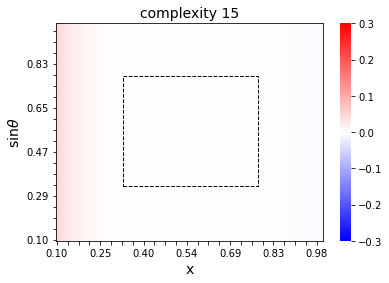}
\caption{Heatmaps in the $(x, \sin\theta)$ plane of the differences between the fit functions found by the symbolic regression and the true target function. The rectangular box marked with a dashed line delineates the domain of values on which the symbolic regression was trained.}\label{fig:error}
\end{figure}

Our results are shown in Table~\ref{tab:Fdirect} and Figure~\ref{fig:loss} in complete analogy to the earlier Table~\ref{tab:Fdirect_2} and Figure~\ref{fig:lossIIIA}. Once again, \pysr~finds the correct expression which is now of complexity 13 (the increase by 2 relative to the result in Table~\ref{tab:Fdirect_2} is due to the numerical prefactor in front of the linear $x$ term). However, the MSE error this time does not go down to machine precision, and instead saturates at around $10^{-4}$, which is due to the statistical uncertainties on the target function values in our training data. Note that ``the knee" in Figure~\ref{fig:loss} is a marker for the true complexity of our target function. 

As mentioned in the introduction, an important principle of explainable AI is ``generalizability", i.e., extrapolating into the region away from the training data. In order to demonstrate this, we repeat the exercise, but this time we train only on the data within the restricted domain shown with the dashed rectangle in Figure~\ref{fig:error}. We then compare the predictions from the fitted functions found by \pysr~to the true target function, by plotting the difference as a heatmap in the $(x, \sin\theta)$ plane (see Figure~6 in \cite{Matchev2022ApJ}). Note that in all four cases, the fit within the training domain is reasonably good, but the extrapolation away from it is successful only for the correct answers at complexities 13 and 15. Furthermore, a careful inspection of the plots in the lower row reveals that the extrapolation is better for complexity 13 compared to complexity 15, even though within the training domain the performance is similar. This fact favors the complexity 13 answer over its competitor.

\subsubsection{Data generated with Madgraph}
\label{parton-madgraph} 

The training data used in the previous Section~\ref{parton-splitting} was generated with the approximate factorized formula (\ref{collinear}) which is valid in the soft/collinear limit. The advantage of doing so is that we knew the answer that we were supposed to get, which allowed us to judge and validate the performance of \pysr. In this subsection, we shall instead generate our training data with a full blown event generator, \amc~\cite{Alwall:2011uj}, which avoids the soft/collinear approximation. For concreteness, we shall use one of the low energy supersymmetry study points from Ref.~\cite{Birkedal:2004xn}, namely, the one with neutralino mass of $M_\chi=225$ GeV. We choose $\sqrt{s} = 500$ GeV at the International Linear Collider. We assumed electromagnetic calorimeter acceptance of $\sin\theta > 0.1$, and required $p_{T\, \gamma} = E_{\gamma} \sin\theta > 7.5$ GeV corresponding to the mask calorimeter acceptance of 1 degree. With that setup, we generated 10 million events as our training data, and repeated the symbolic regression exercise with default \pysr~parameters.

\begin{table}[t!]
\centering
\scalebox{0.89}{
\renewcommand\arraystretch{1.8}
\begin{tabular}{||c|c|c|c||}
\hline
Complexity & Fitted function & MSE & Score \\
\hline
\hline
5 & $17.07-98.85x$ & $1.968$ & $0.671$ \\ [1mm]
7 & $17.08-98.85x+x^2$ & $1.968$ & $1.6\times10^{-5}$ \\ [1mm]
9 & $-11.72+\dfrac{2.42-0.057/x}{x}$ & $0.115$ & $1.419$ \\ [2mm]
11 & $x-11.97+\dfrac{2.44-0.057/x}{x}$ &$0.113$ & $0.007$ \\ [2mm]
13 &$2x-12.23+\dfrac{2.46-0.058/x}{x}$  & $0.112$ & $0.007$ \\ [2mm]
15 &$3x-12.48+\dfrac{2.48+0.058/x}{x}$  & $0.111$ & $0.006$\\ [2mm]
\hline\hline
\end{tabular}
}
\caption{The same as Table~\ref{tab:Fdirect}, but for the symbolic regression exercise with \amc~training data performed in Section~\ref{parton-madgraph}. 
\label{tab:Gen_NLO}}
\end{table}

\begin{figure}[t]
    \centering
    \includegraphics[width=0.47\textwidth]{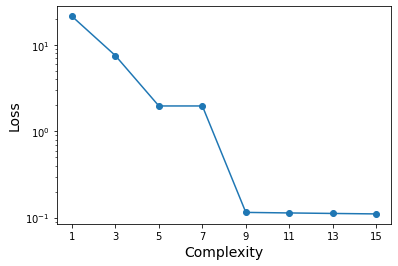}
    \caption{The same as Figure~\ref{fig:loss}, but for the symbolic regression exercise with \amc~ training data performed in Section~\ref{parton-madgraph}. 
    \label{fig:lossMG5} }
\end{figure}

\begin{figure*}[t]
    \centering
    \includegraphics[width=0.45\textwidth]{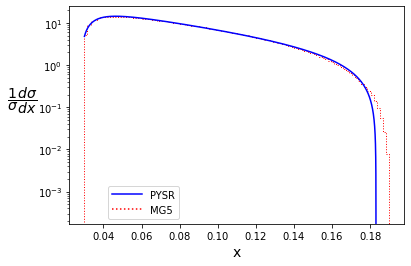}
    \includegraphics[width=0.45\textwidth]{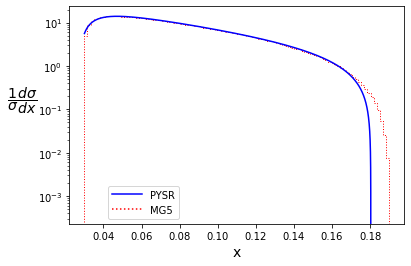}
    \caption{    
    Unit-normalized distribution of the events in the training data (red) and PYSR output (blue). The results in the left panel are from Section~\ref{parton-madgraph} and do not include detector effects, while the results in the right panel are from Section~\ref{detector-madgraph} and account for the detector resolution. \label{fig:distMG5} }
\end{figure*}

In analogy to the earlier Tables~\ref{tab:Fdirect_2} and \ref{tab:Fdirect} and Figures~\ref{fig:lossIIIA} and \ref{fig:loss}, we present the results in Table~\ref{tab:Gen_NLO} and Figure~\ref{fig:lossMG5}, where for simplicity we focus on the $x$-dependence only. The knee in Figure~\ref{fig:lossMG5} is observed at complexity 9, which also has the highest score in Table~\ref{tab:Gen_NLO}. The form of the function resembles that of (\ref{llog}), but the coefficients are modified. The expressions at higher complexities (11, 13 and 15), while having comparable MSE, might be disfavored using the method discussed at the end of the previous subsection, see Figure~\ref{fig:error}.

Since in this example we do not have a simple analytical answer as a point of reference, the only way to judge the quality of the answer is to numerically compare to the distribution in the training data. In Figure~\ref{fig:distMG5}, we show the unit-normalized distribution of the events in the training data (red) and PYSR output (blue). The results from the current subsection are shown in the left panel, where the blue line corresponds to the fitted function at complexity 9. We see that the symbolic regression was capable of producing a simple analytical expression which describes the data quite well, the main visible discrepancy is in the low statistics tail which is not represented well in the training data, and furthermore, is not relevant for the experimental analysis.

\subsection{Learning from detector-level MC data}
\label{sec:F_MC_detector}

So far in this section we have been ignoring any instrumental effects, so that the observed distribution followed the theoretical formula (up to statistical errors). In this section we shall add the effects of the detector resolution which would in principle cause the result from the symbolic regression to differ slightly from the theoretical prediction at gen-level.

\subsubsection{Data generated using a splitting function}
\label{detector-splitting} 

\begin{table*}[t]
\centering
\scalebox{0.86}{
\renewcommand\arraystretch{1.7}
\begin{tabular}{||c|c|c|c||}
\hline
Complexity & Fitted function & MSE & Score \\
\hline
\hline
\multicolumn{4}{||c||}{$\sigma=0.01$}\\
\hline
9 & $ (-0.039+0.064/x)/\sin^2\theta$ & $8.41\times10^{-3}$ & $1.58$ \\
11 & $ 0.056/(x\sin^2\theta(x+0.82))$ &$5.91\times10^{-4}$ & $1.33$ \\
13 &$ (-0.068+0.068/x+0.034x)/\sin^2\theta$  & $2.46\times10^{-4}$ & $0.438$ \\
17 &$ [(-0.067+0.067/x+0.034x+\sin\theta)/\sin\theta-1.00]/\sin\theta$  & $2.44\times10^{-4}$ & $2.11\times10^{-3}$\\
\hline\hline
\multicolumn{4}{||c||}{$\sigma = 0.03$}\\
\hline
9 & $ (-0.039+0.064/x)/\sin^2\theta$ & $8.71\times10^{-3}$ & $1.57$ \\
11 & $ 0.056/(x\sin^2\theta(x+0.81))$ &$7.95\times10^{-4}$ & $1.20$ \\
13 &$ (-0.068+0.068/x+0.034x)/\sin^2\theta$  & $5.33\times10^{-4}$ & $0.20$ \\
15 &$ (-0.068+0.068/x+0.034x)/\sin^2\theta-1.24\times10^{-3}$  & $5.32\times10^{-4}$ & $1.06\times10^{-3}$\\
\hline\hline
\multicolumn{4}{||c||}{$\sigma = 0.05$}\\
\hline
9 & $ (-0.039+0.064/x)/\sin^2\theta$ & $1.16\times10^{-2}$ & $1.44$ \\
11 & $ 0.056/(x\sin^2\theta(x+0.81))$ &$3.89\times10^{-3}$ & $5.46\times10^{-1}$ \\
13 &$ (-0.067+0.068/x+0.033x)/\sin^2\theta$  & $3.70\times10^{-3}$ & $2.51\times10^{-2}$ \\
15 &$ [(-0.067+0.068/x+0.033x)/\sin\theta-1.10\times10^{-3}]/\sin\theta$  & $3.70\times10^{-3}$ & $2.78\times10^{-4}$\\
\hline\hline
\multicolumn{4}{||c||}{$\sigma = 0.1$}\\
\hline
9 & $ (-0.039+0.064/x)/\sin^2\theta$ & $3.90\times10^{-2}$ & $9.00\times10^{-1}$ \\
11 & $ 0.056/(x\sin^2\theta(x+0.82))$ &$3.30\times10^{-2}$ & $8.28\times10^{-2}$ \\
13 &$ (-0.064+0.067/x+0.029x)/\sin^2\theta$  & $3.27\times10^{-2}$ & $3.60\times10^{-3}$ \\
15 &$ 0.078/[\sin^2\theta(1.62x^2+x+0.014)]$  & $3.17\times10^{-2}$ & $1.75\times10^{-2}$\\
\hline\hline
\end{tabular} }
\caption{Results from the symbolic regression exercise performed in Section~\ref{detector-splitting} for several values of the detector resolution parameter $\sigma$: 0.01, 0.03, 0.05 and 0.10. 
\label{tab:F_MC}
}
\end{table*}

Here we repeat the exercise from Section~\ref{parton-splitting}, but account for the detector resolution via Gaussian smearing of the energy (but not direction) of the photon with some resolution parameter $\sigma$. By varying the value of $\sigma$, we shall investigate the impact of the detector on our results, which are collected in Table~\ref{tab:F_MC} for four different values of $\sigma$: 0.01, 0.03, 0.05 and 0.10. In all cases, we observe the expected $\sin^{-2}\theta$ dependence. We note that when the detector effects are relatively mild, $\sigma\lesssim  5\%$, the $x$ dependence is well recovered as well. This fact --- that symbolic regression appears to be robust against noise --- has been observed in other independent studies as well \cite{DBLP:journals/corr/abs-1912-04871}.

\begin{figure}[t!]
    \centering
    \includegraphics[width=0.235\textwidth]{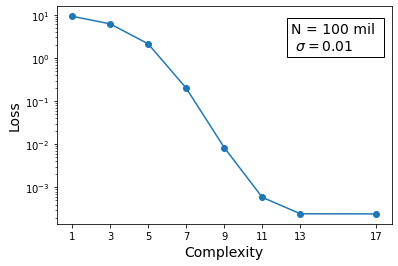}
    \includegraphics[width=0.235\textwidth]{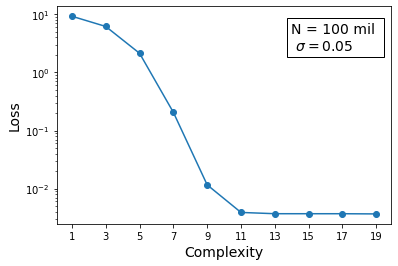} \\
    \includegraphics[width=0.235\textwidth]{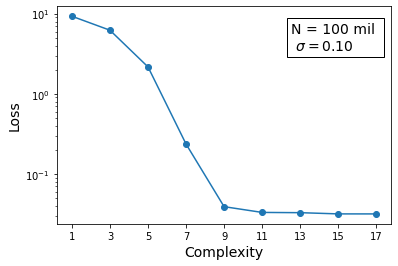}
    \includegraphics[width=0.235\textwidth]{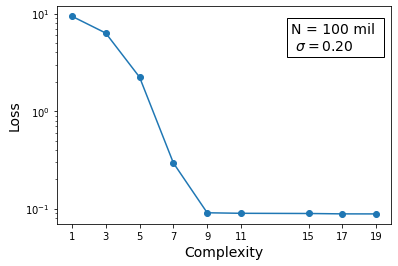}
\caption{The same as Figure \ref{fig:loss}, but for the exercise performed in Section~\ref{detector-splitting} with the added detector smearing. Results are shown for several values of the detector resolution parameter $\sigma$ as labelled in the plots. }
    \label{fig:loss_smearing}
\end{figure}

\begin{figure}[t]
    \centering
    \includegraphics[width=0.4\textwidth]{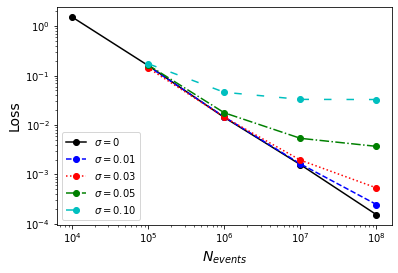}
    \caption{Loss as a function of the number of events in the training data, for several values of the detector resolution parameter $\sigma$. In each case, we chose to show the \pysr~result whose complexity is at the ``knee" of the corresponding plot from Figure~\ref{fig:loss_smearing}.  \label{fig:loss2}}
\end{figure}

In analogy to Figure~\ref{fig:loss}, in Figure~\ref{fig:loss_smearing} we show the evolution of the MSE loss with the complexity of the fitted function, for several different values of $\sigma$: 0.01, 0.05, 0.10 and 0.20. The ``knee" structure is again evident, and the location of the knee depends slightly on the amount of applied smearing.

Since the exercises in this subsection include both errors due to the finite statistics and due to the detector resolution, it is instructive to look at the interplay of the two types of errors as a function of the number of events $N_{\rm events}$ in the training data, see Figure~\ref{fig:loss2}. When the detector effects are absent ($\sigma=0$, black line), the average loss improves as $N_{\rm events}$ increases, since statistical errors scale as $1/\sqrt{N_{\rm events}}$. On the other hand, the detector effects are not influenced by $N_{\rm events}$, and at some point will start to dominate the error budget. As a result, as illustrated in Figure~\ref{fig:loss2}, the MSE loss will start to deviate from the benchmark case of $\sigma=0$. The exact point where this deviation occurs, depends on the size of the detector smearing parameter --- the larger the smearing, the earlier the loss saturates.

\subsubsection{Data generated with MadGraph}
\label{detector-madgraph} 

\begin{figure}[t]
    \centering
    \includegraphics[width=0.47\textwidth]{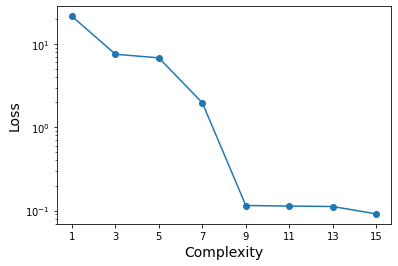}
    \caption{
    The same as Figure \ref{fig:lossMG5}, but adding the effects of the calorimeter resolution as in Section~\ref{detector-madgraph}.}\label{fig:lossMG5smeared}
\end{figure}

Finally, we repeat the exercise from Section \ref{parton-madgraph} with the addition of calorimeter detector resolution typical of the ILC, $\delta E/E = 0.17/\sqrt{E}$ \cite{CALICE:2008kht,Bambade:2019fyw,Habermehl:2020njb,ILDConceptGroup:2020sfq}. The results are displayed in Table~\ref{tab:Gen_NLO_detector} and Figure~\ref{fig:lossMG5smeared}, which are the analogues of Table~\ref{tab:Gen_NLO} and Figure~\ref{fig:lossMG5} from Section~\ref{parton-madgraph}.
The results are as expected, based on what we have observed in the previous subsections. The corresponding predicted differential distribution is shown in the right panel of Figure~\ref{fig:distMG5}.

\begin{table}[t]
\centering
\renewcommand\arraystretch{1.8}
\scalebox{0.8}{
\begin{tabular}{||c|c|c|c||}
\hline
Complexity & Fitted function & MSE & Score \\
\hline
\hline
5 & $0.724/(x+0.015)$ & $6.82$ & $0.051$ \\[2mm]
7 & $18.07+  98.855446x$ & $1.97$ & $0.623$ \\[2mm]
9 & $-11.72+\dfrac{2.42-0.057/x}{x}$ & $0.115$ & $1.419$ \\[2mm]
11 & $x-11.97+\dfrac{2.44-0.057/x}{x}$ &$0.114$ & $0.007$ \\[2mm]
13 &$2x-12.23+\dfrac{2.46-0.058/x}{x}$  & $0.112$ & $0.007$ \\[2mm]
15 &$-x-9.90+\dfrac{2.04+(-0.03-0.0004/x)/x}{x}$  & $0.091$ & $0.102$\\[2mm]
\hline\hline
\end{tabular} }
\caption{Results from the symbolic regression exercise performed in Section \ref{detector-madgraph} including detector effects in the training data.
\label{tab:Gen_NLO_detector}}
\end{table}

\section{Conclusions and outlook}
\label{sec:conclusions}

This study adds to the already wide range of applications of modern machine learning to event generation and simulation-based inference in collider phenomenology \cite{Butter:2022rso}. We demonstrated the use of symbolic regression for two common problems in high energy particle physics. First, in the case of kinematic or event variables which are defined through some kind of an algorithm, the symbolic regression produces analytical formulas whose accuracy is limited only by the desired functional complexity. In Section~\ref{sec:mt2} we showed how to do this in the example of the stransverse mass variable $M_{T2}$ --- we were able to rederive all known analytical formulas for $M_{T2}$ in certain special transverse momentum configurations. Second, the symbolic regression can also produce analytical formulas for certain kinematic distributions of interest, for which theoretical results are unknown or difficult to obtain. In fact, parametrizing the observed distributions in the data with analytical formulas is a standard task in many analyses which attempt to measure the background from data. In Section~\ref{sec:F} we demonstrated that this fit can be done either at the gen-level (Section~\ref{sec:F_MC}) or at the detector level (Section~\ref{sec:F_MC_detector}). Note that this last exercise is a non-trivial result, which involves the convolution of the parton-level analytical result with the transfer function describing the detector. To the best of our knowledge, such analytical expressions are rarely discussed in the literature.

The work presented here can be extended in several directions. For example, the $M_{T2}$ concept can be readily applied to more complex event topologies, where one has several choices of designating parent and daughter particles, leading to a menagerie of different ``subsystem" $M_{T2}$ variables \cite{Kawagoe:2004rz,Burns:2008va}. It would be interesting to see whether the symbolic regression can ``derive" the correct answer for $M_{T2}$ in the general case, for which no analytical formula is known. One could also explore other modern techniques for symbolic regression that are adaptable to high dimensional data \cite{Arechiga2021,https://doi.org/10.48550/arxiv.2201.04600,https://doi.org/10.48550/arxiv.2204.10532,https://doi.org/10.48550/arxiv.2205.11798}.

\acknowledgements
We thank A.~Roman for collaboration in the early stages of this work. K. Kong and K. Matchev would like to thank the Aspen Center for Physics for hospitality during the completion of this work, supported in part by National Science Foundation grant PHY-1607611. This work is supported in parts by US DE-SC0019474 and DOE DE-SC0022148.

\bibliography{refs}

\begin{thebibliography}{65}%
\makeatletter
\providecommand \@ifxundefined [1]{%
 \@ifx{#1\undefined}
}%
\providecommand \@ifnum [1]{%
 \ifnum #1\expandafter \@firstoftwo
 \else \expandafter \@secondoftwo
 \fi
}%
\providecommand \@ifx [1]{%
 \ifx #1\expandafter \@firstoftwo
 \else \expandafter \@secondoftwo
 \fi
}%
\providecommand \natexlab [1]{#1}%
\providecommand \enquote  [1]{``#1''}%
\providecommand \bibnamefont  [1]{#1}%
\providecommand \bibfnamefont [1]{#1}%
\providecommand \citenamefont [1]{#1}%
\providecommand \href@noop [0]{\@secondoftwo}%
\providecommand \href [0]{\begingroup \@sanitize@url \@href}%
\providecommand \@href[1]{\@@startlink{#1}\@@href}%
\providecommand \@@href[1]{\endgroup#1\@@endlink}%
\providecommand \@sanitize@url [0]{\catcode `\\12\catcode `\$12\catcode
  `\&12\catcode `\#12\catcode `\^12\catcode `\_12\catcode `\%12\relax}%
\providecommand \@@startlink[1]{}%
\providecommand \@@endlink[0]{}%
\providecommand \url  [0]{\begingroup\@sanitize@url \@url }%
\providecommand \@url [1]{\endgroup\@href {#1}{\urlprefix }}%
\providecommand \urlprefix  [0]{URL }%
\providecommand \Eprint [0]{\href }%
\providecommand \doibase [0]{http://dx.doi.org/}%
\providecommand \selectlanguage [0]{\@gobble}%
\providecommand \bibinfo  [0]{\@secondoftwo}%
\providecommand \bibfield  [0]{\@secondoftwo}%
\providecommand \translation [1]{[#1]}%
\providecommand \BibitemOpen [0]{}%
\providecommand \bibitemStop [0]{}%
\providecommand \bibitemNoStop [0]{.\EOS\space}%
\providecommand \EOS [0]{\spacefactor3000\relax}%
\providecommand \BibitemShut  [1]{\csname bibitem#1\endcsname}%
\let\auto@bib@innerbib\@empty
\bibitem [{\citenamefont {Langley}(1977)}]{Langley1977}%
  \BibitemOpen
  \bibfield  {author} {\bibinfo {author} {\bibfnamefont {Pat}\ \bibnamefont
  {Langley}},\ }\bibfield  {title} {\enquote {\bibinfo {title} {Bacon: A
  production system that discovers empirical laws},}\ }in\ \href@noop {} {\emph
  {\bibinfo {booktitle} {IJCAI}}}\ (\bibinfo {year} {1977})\BibitemShut
  {NoStop}%
\bibitem [{\citenamefont {Langley}\ \emph {et~al.}(1987)\citenamefont
  {Langley}, \citenamefont {Simon},\ and\ \citenamefont
  {Bradshaw}}]{Langley1987}%
  \BibitemOpen
  \bibfield  {author} {\bibinfo {author} {\bibfnamefont {Pat}\ \bibnamefont
  {Langley}}, \bibinfo {author} {\bibfnamefont {Herbert~A.}\ \bibnamefont
  {Simon}}, \ and\ \bibinfo {author} {\bibfnamefont {Gary~L.}\ \bibnamefont
  {Bradshaw}},\ }\enquote {\bibinfo {title} {Heuristics for empirical
  discovery},}\ in\ \href {\doibase 10.1007/978-3-642-82742-6_2} {\emph
  {\bibinfo {booktitle} {Computational Models of Learning}}},\ \bibinfo
  {editor} {edited by\ \bibinfo {editor} {\bibfnamefont {Leonard}\ \bibnamefont
  {Bolc}}}\ (\bibinfo  {publisher} {Springer Berlin Heidelberg},\ \bibinfo
  {address} {Berlin, Heidelberg},\ \bibinfo {year} {1987})\ pp.\ \bibinfo
  {pages} {21--54}\BibitemShut {NoStop}%
\bibitem [{\citenamefont {Kokar}(1986)}]{Kokar1986}%
  \BibitemOpen
  \bibfield  {author} {\bibinfo {author} {\bibfnamefont {Mieczyslaw}\
  \bibnamefont {Kokar}},\ }\bibfield  {title} {\enquote {\bibinfo {title}
  {Determining arguments of invariant functional descriptions.}}\ }\href
  {\doibase 10.1023/A:1022818816206} {\bibfield  {journal} {\bibinfo  {journal}
  {Machine Learning}\ }\textbf {\bibinfo {volume} {1}},\ \bibinfo {pages}
  {403--422} (\bibinfo {year} {1986})}\BibitemShut {NoStop}%
\bibitem [{\citenamefont {Langley}\ and\ \citenamefont
  {Zytkow}(1989)}]{Langley1989}%
  \BibitemOpen
  \bibfield  {author} {\bibinfo {author} {\bibfnamefont {Pat}\ \bibnamefont
  {Langley}}\ and\ \bibinfo {author} {\bibfnamefont {Jan~M.}\ \bibnamefont
  {Zytkow}},\ }\bibfield  {title} {\enquote {\bibinfo {title} {Data-driven
  approaches to empirical discovery},}\ }\href {\doibase
  https://doi.org/10.1016/0004-3702(89)90051-9} {\bibfield  {journal} {\bibinfo
   {journal} {Artificial Intelligence}\ }\textbf {\bibinfo {volume} {40}},\
  \bibinfo {pages} {283--312} (\bibinfo {year} {1989})}\BibitemShut {NoStop}%
\bibitem [{\citenamefont {Zembowicz}\ and\ \citenamefont
  {\.{Z}ytkow}(1992)}]{Zembowicz1992}%
  \BibitemOpen
  \bibfield  {author} {\bibinfo {author} {\bibfnamefont {Robert}\ \bibnamefont
  {Zembowicz}}\ and\ \bibinfo {author} {\bibfnamefont {Jan~M.}\ \bibnamefont
  {\.{Z}ytkow}},\ }\bibfield  {title} {\enquote {\bibinfo {title} {Discovery of
  equations: Experimental evaluation of convergence},}\ }in\ \href@noop {}
  {\emph {\bibinfo {booktitle} {Proceedings of the Tenth National Conference on
  Artificial Intelligence}}},\ \bibinfo {series and number} {AAAI'92}\
  (\bibinfo  {publisher} {AAAI Press},\ \bibinfo {year} {1992})\ p.\ \bibinfo
  {pages} {70–75}\BibitemShut {NoStop}%
\bibitem [{\citenamefont {Todorovski}\ and\ \citenamefont
  {Dzeroski}(1997)}]{Todorovski1997}%
  \BibitemOpen
  \bibfield  {author} {\bibinfo {author} {\bibfnamefont {Ljupco}\ \bibnamefont
  {Todorovski}}\ and\ \bibinfo {author} {\bibfnamefont {Saso}\ \bibnamefont
  {Dzeroski}},\ }\bibfield  {title} {\enquote {\bibinfo {title} {Declarative
  bias in equation discovery},}\ }in\ \href@noop {} {\emph {\bibinfo
  {booktitle} {Proceedings of the Fourteenth International Conference on
  Machine Learning}}}\ (\bibinfo  {publisher} {Morgan Kaufmann},\ \bibinfo
  {year} {1997})\ pp.\ \bibinfo {pages} {376--384}\BibitemShut {NoStop}%
\bibitem [{\citenamefont {{Bongard}}\ and\ \citenamefont
  {{Lipson}}(2007)}]{Bongard2007}%
  \BibitemOpen
  \bibfield  {author} {\bibinfo {author} {\bibfnamefont {Josh}\ \bibnamefont
  {{Bongard}}}\ and\ \bibinfo {author} {\bibfnamefont {Hod}\ \bibnamefont
  {{Lipson}}},\ }\bibfield  {title} {\enquote {\bibinfo {title} {{From the
  Cover: Automated reverse engineering of nonlinear dynamical systems}},}\
  }\href {\doibase 10.1073/pnas.0609476104} {\bibfield  {journal} {\bibinfo
  {journal} {Proceedings of the National Academy of Science}\ }\textbf
  {\bibinfo {volume} {104}},\ \bibinfo {pages} {9943--9948} (\bibinfo {year}
  {2007})}\BibitemShut {NoStop}%
\bibitem [{\citenamefont {{Schmidt}}\ and\ \citenamefont
  {{Lipson}}(2009)}]{Schmidt2009}%
  \BibitemOpen
  \bibfield  {author} {\bibinfo {author} {\bibfnamefont {Michael}\ \bibnamefont
  {{Schmidt}}}\ and\ \bibinfo {author} {\bibfnamefont {Hod}\ \bibnamefont
  {{Lipson}}},\ }\bibfield  {title} {\enquote {\bibinfo {title} {{Distilling
  Free-Form Natural Laws from Experimental Data}},}\ }\href {\doibase
  10.1126/science.1165893} {\bibfield  {journal} {\bibinfo  {journal}
  {Science}\ }\textbf {\bibinfo {volume} {324}},\ \bibinfo {pages} {81}
  (\bibinfo {year} {2009})}\BibitemShut {NoStop}%
\bibitem [{\citenamefont {{Battaglia}}\ \emph {et~al.}(2016)\citenamefont
  {{Battaglia}}, \citenamefont {{Pascanu}}, \citenamefont {{Lai}},
  \citenamefont {{Rezende}},\ and\ \citenamefont
  {{Kavukcuoglu}}}]{Battaglia2016}%
  \BibitemOpen
  \bibfield  {author} {\bibinfo {author} {\bibfnamefont {Peter~W.}\
  \bibnamefont {{Battaglia}}}, \bibinfo {author} {\bibfnamefont {Razvan}\
  \bibnamefont {{Pascanu}}}, \bibinfo {author} {\bibfnamefont {Matthew}\
  \bibnamefont {{Lai}}}, \bibinfo {author} {\bibfnamefont {Danilo}\
  \bibnamefont {{Rezende}}}, \ and\ \bibinfo {author} {\bibfnamefont {Koray}\
  \bibnamefont {{Kavukcuoglu}}},\ }\bibfield  {title} {\enquote {\bibinfo
  {title} {{Interaction Networks for Learning about Objects, Relations and
  Physics}},}\ }\href@noop {} {\bibfield  {journal} {\bibinfo  {journal} {arXiv
  e-prints}\ ,\ \bibinfo {eid} {arXiv:1612.00222}} (\bibinfo {year} {2016})},\
  \Eprint {http://arxiv.org/abs/1612.00222} {arXiv:1612.00222 [cs.AI]}
  \BibitemShut {NoStop}%
\bibitem [{\citenamefont {{Chang}}\ \emph {et~al.}(2016)\citenamefont
  {{Chang}}, \citenamefont {{Ullman}}, \citenamefont {{Torralba}},\ and\
  \citenamefont {{Tenenbaum}}}]{Chang2016}%
  \BibitemOpen
  \bibfield  {author} {\bibinfo {author} {\bibfnamefont {Michael~B.}\
  \bibnamefont {{Chang}}}, \bibinfo {author} {\bibfnamefont {Tomer}\
  \bibnamefont {{Ullman}}}, \bibinfo {author} {\bibfnamefont {Antonio}\
  \bibnamefont {{Torralba}}}, \ and\ \bibinfo {author} {\bibfnamefont
  {Joshua~B.}\ \bibnamefont {{Tenenbaum}}},\ }\bibfield  {title} {\enquote
  {\bibinfo {title} {{A Compositional Object-Based Approach to Learning
  Physical Dynamics}},}\ }\href@noop {} {\bibfield  {journal} {\bibinfo
  {journal} {arXiv e-prints}\ ,\ \bibinfo {eid} {arXiv:1612.00341}} (\bibinfo
  {year} {2016})},\ \Eprint {http://arxiv.org/abs/1612.00341} {arXiv:1612.00341
  [cs.AI]} \BibitemShut {NoStop}%
\bibitem [{\citenamefont {Guimerà}\ \emph {et~al.}(2020)\citenamefont
  {Guimerà}, \citenamefont {Reichardt}, \citenamefont {Aguilar-Mogas},
  \citenamefont {Massucci}, \citenamefont {Miranda}, \citenamefont
  {Pallarès},\ and\ \citenamefont {Sales-Pardo}}]{Guimera2020}%
  \BibitemOpen
  \bibfield  {author} {\bibinfo {author} {\bibfnamefont {Roger}\ \bibnamefont
  {Guimerà}}, \bibinfo {author} {\bibfnamefont {Ignasi}\ \bibnamefont
  {Reichardt}}, \bibinfo {author} {\bibfnamefont {Antoni}\ \bibnamefont
  {Aguilar-Mogas}}, \bibinfo {author} {\bibfnamefont {Francesco~A.}\
  \bibnamefont {Massucci}}, \bibinfo {author} {\bibfnamefont {Manuel}\
  \bibnamefont {Miranda}}, \bibinfo {author} {\bibfnamefont {Jordi}\
  \bibnamefont {Pallarès}}, \ and\ \bibinfo {author} {\bibfnamefont {Marta}\
  \bibnamefont {Sales-Pardo}},\ }\bibfield  {title} {\enquote {\bibinfo {title}
  {A bayesian machine scientist to aid in the solution of challenging
  scientific problems},}\ }\href {\doibase 10.1126/sciadv.aav6971} {\bibfield
  {journal} {\bibinfo  {journal} {Science Advances}\ }\textbf {\bibinfo
  {volume} {6}},\ \bibinfo {pages} {eaav6971} (\bibinfo {year} {2020})},\
  \Eprint
  {http://arxiv.org/abs/https://www.science.org/doi/pdf/10.1126/sciadv.aav6971}
  {https://www.science.org/doi/pdf/10.1126/sciadv.aav6971} \BibitemShut
  {NoStop}%
\bibitem [{\citenamefont {Udrescu}\ and\ \citenamefont
  {Tegmark}(2020)}]{Udrescu:2019mnk}%
  \BibitemOpen
  \bibfield  {author} {\bibinfo {author} {\bibfnamefont {Silviu-Marian}\
  \bibnamefont {Udrescu}}\ and\ \bibinfo {author} {\bibfnamefont {Max}\
  \bibnamefont {Tegmark}},\ }\bibfield  {title} {\enquote {\bibinfo {title}
  {{AI Feynman: a Physics-Inspired Method for Symbolic Regression}},}\ }\href
  {\doibase 10.1126/sciadv.aay2631} {\bibfield  {journal} {\bibinfo  {journal}
  {Sci. Adv.}\ }\textbf {\bibinfo {volume} {6}},\ \bibinfo {pages} {eaay2631}
  (\bibinfo {year} {2020})},\ \Eprint {http://arxiv.org/abs/1905.11481}
  {arXiv:1905.11481 [physics.comp-ph]} \BibitemShut {NoStop}%
\bibitem [{\citenamefont {Cranmer}\ \emph {et~al.}(2020)\citenamefont
  {Cranmer}, \citenamefont {Sanchez-Gonzalez}, \citenamefont {Battaglia},
  \citenamefont {Xu}, \citenamefont {Cranmer}, \citenamefont {Spergel},\ and\
  \citenamefont {Ho}}]{Cranmer:2020wew}%
  \BibitemOpen
  \bibfield  {author} {\bibinfo {author} {\bibfnamefont {Miles}\ \bibnamefont
  {Cranmer}}, \bibinfo {author} {\bibfnamefont {Alvaro}\ \bibnamefont
  {Sanchez-Gonzalez}}, \bibinfo {author} {\bibfnamefont {Peter}\ \bibnamefont
  {Battaglia}}, \bibinfo {author} {\bibfnamefont {Rui}\ \bibnamefont {Xu}},
  \bibinfo {author} {\bibfnamefont {Kyle}\ \bibnamefont {Cranmer}}, \bibinfo
  {author} {\bibfnamefont {David}\ \bibnamefont {Spergel}}, \ and\ \bibinfo
  {author} {\bibfnamefont {Shirley}\ \bibnamefont {Ho}},\ }\bibfield  {title}
  {\enquote {\bibinfo {title} {{Discovering Symbolic Models from Deep Learning
  with Inductive Biases}},}\ }\href@noop {} {\bibfield  {journal} {\bibinfo
  {journal} {NeurIPS2020}\ } (\bibinfo {year} {2020})},\ \Eprint
  {http://arxiv.org/abs/2006.11287} {arXiv:2006.11287 [cs.LG]} \BibitemShut
  {NoStop}%
\bibitem [{\citenamefont {Liu}\ \emph {et~al.}(2022)\citenamefont {Liu},
  \citenamefont {Madhavan},\ and\ \citenamefont {Tegmark}}]{liu2022ai}%
  \BibitemOpen
  \bibfield  {author} {\bibinfo {author} {\bibfnamefont {Ziming}\ \bibnamefont
  {Liu}}, \bibinfo {author} {\bibfnamefont {Varun}\ \bibnamefont {Madhavan}}, \
  and\ \bibinfo {author} {\bibfnamefont {Max}\ \bibnamefont {Tegmark}},\
  }\href@noop {} {\enquote {\bibinfo {title} {Ai poincar\'{e} 2.0: Machine
  learning conservation laws from differential equations},}\ } (\bibinfo {year}
  {2022}),\ \Eprint {http://arxiv.org/abs/2203.12610} {arXiv:2203.12610
  [cs.LG]} \BibitemShut {NoStop}%
\bibitem [{\citenamefont {Matsubara}\ \emph {et~al.}(2022)\citenamefont
  {Matsubara}, \citenamefont {Chiba}, \citenamefont {Igarashi}, \citenamefont
  {Taniai},\ and\ \citenamefont
  {Ushiku}}]{https://doi.org/10.48550/arxiv.2206.10540}%
  \BibitemOpen
  \bibfield  {author} {\bibinfo {author} {\bibfnamefont {Yoshitomo}\
  \bibnamefont {Matsubara}}, \bibinfo {author} {\bibfnamefont {Naoya}\
  \bibnamefont {Chiba}}, \bibinfo {author} {\bibfnamefont {Ryo}\ \bibnamefont
  {Igarashi}}, \bibinfo {author} {\bibfnamefont {Tatsunori}\ \bibnamefont
  {Taniai}}, \ and\ \bibinfo {author} {\bibfnamefont {Yoshitaka}\ \bibnamefont
  {Ushiku}},\ }\href {\doibase 10.48550/ARXIV.2206.10540} {\enquote {\bibinfo
  {title} {Rethinking symbolic regression datasets and benchmarks for
  scientific discovery},}\ } (\bibinfo {year} {2022}),\ \Eprint
  {http://arxiv.org/abs/2206.10540} {arXiv:2206.10540 [cs.LG]} \BibitemShut
  {NoStop}%
\bibitem [{\citenamefont {Cranmer}\ \emph {et~al.}(2019)\citenamefont
  {Cranmer}, \citenamefont {Xu}, \citenamefont {Battaglia},\ and\ \citenamefont
  {Ho}}]{Cranmer2019}%
  \BibitemOpen
  \bibfield  {author} {\bibinfo {author} {\bibfnamefont {Miles~D.}\
  \bibnamefont {Cranmer}}, \bibinfo {author} {\bibfnamefont {Rui}\ \bibnamefont
  {Xu}}, \bibinfo {author} {\bibfnamefont {Peter}\ \bibnamefont {Battaglia}}, \
  and\ \bibinfo {author} {\bibfnamefont {Shirley}\ \bibnamefont {Ho}},\ }\href
  {\doibase 10.48550/ARXIV.1909.05862} {\enquote {\bibinfo {title} {Learning
  symbolic physics with graph networks},}\ } (\bibinfo {year} {2019}),\ \Eprint
  {http://arxiv.org/abs/1909.05862} {arXiv:1909.05862 [cs.LG]} \BibitemShut
  {NoStop}%
\bibitem [{\citenamefont {{Delgado}}\ \emph {et~al.}(2021)\citenamefont
  {{Delgado}}, \citenamefont {{Wadekar}}, \citenamefont {{Hadzhiyska}},
  \citenamefont {{Bose}}, \citenamefont {{Hernquist}},\ and\ \citenamefont
  {{Ho}}}]{2021arXiv211102422D}%
  \BibitemOpen
  \bibfield  {author} {\bibinfo {author} {\bibfnamefont {Ana~Maria}\
  \bibnamefont {{Delgado}}}, \bibinfo {author} {\bibfnamefont {Digvijay}\
  \bibnamefont {{Wadekar}}}, \bibinfo {author} {\bibfnamefont {Boryana}\
  \bibnamefont {{Hadzhiyska}}}, \bibinfo {author} {\bibfnamefont {Sownak}\
  \bibnamefont {{Bose}}}, \bibinfo {author} {\bibfnamefont {Lars}\ \bibnamefont
  {{Hernquist}}}, \ and\ \bibinfo {author} {\bibfnamefont {Shirley}\
  \bibnamefont {{Ho}}},\ }\bibfield  {title} {\enquote {\bibinfo {title}
  {{Modeling the galaxy-halo connection with machine learning}},}\ }\href@noop
  {} {\bibfield  {journal} {\bibinfo  {journal} {arXiv e-prints}\ ,\ \bibinfo
  {eid} {arXiv:2111.02422}} (\bibinfo {year} {2021})},\ \Eprint
  {http://arxiv.org/abs/2111.02422} {arXiv:2111.02422 [astro-ph.CO]}
  \BibitemShut {NoStop}%
\bibitem [{\citenamefont {{Iten}}\ \emph {et~al.}(2020)\citenamefont {{Iten}},
  \citenamefont {{Metger}}, \citenamefont {{Wilming}}, \citenamefont {{del
  Rio}},\ and\ \citenamefont {{Renner}}}]{Iten2020}%
  \BibitemOpen
  \bibfield  {author} {\bibinfo {author} {\bibfnamefont {Raban}\ \bibnamefont
  {{Iten}}}, \bibinfo {author} {\bibfnamefont {Tony}\ \bibnamefont {{Metger}}},
  \bibinfo {author} {\bibfnamefont {Henrik}\ \bibnamefont {{Wilming}}},
  \bibinfo {author} {\bibfnamefont {L{\'\i}dia}\ \bibnamefont {{del Rio}}}, \
  and\ \bibinfo {author} {\bibfnamefont {Renato}\ \bibnamefont {{Renner}}},\
  }\bibfield  {title} {\enquote {\bibinfo {title} {{Discovering Physical
  Concepts with Neural Networks}},}\ }\href {\doibase
  10.1103/PhysRevLett.124.010508} {\bibfield  {journal} {\bibinfo  {journal}
  {\prl}\ }\textbf {\bibinfo {volume} {124}},\ \bibinfo {eid} {010508}
  (\bibinfo {year} {2020})},\ \Eprint {http://arxiv.org/abs/1807.10300}
  {arXiv:1807.10300 [quant-ph]} \BibitemShut {NoStop}%
\bibitem [{\citenamefont {{Lemos}}\ \emph {et~al.}(2022)\citenamefont
  {{Lemos}}, \citenamefont {{Jeffrey}}, \citenamefont {{Cranmer}},
  \citenamefont {{Ho}},\ and\ \citenamefont {{Battaglia}}}]{Lemos2022}%
  \BibitemOpen
  \bibfield  {author} {\bibinfo {author} {\bibfnamefont {Pablo}\ \bibnamefont
  {{Lemos}}}, \bibinfo {author} {\bibfnamefont {Niall}\ \bibnamefont
  {{Jeffrey}}}, \bibinfo {author} {\bibfnamefont {Miles}\ \bibnamefont
  {{Cranmer}}}, \bibinfo {author} {\bibfnamefont {Shirley}\ \bibnamefont
  {{Ho}}}, \ and\ \bibinfo {author} {\bibfnamefont {Peter}\ \bibnamefont
  {{Battaglia}}},\ }\bibfield  {title} {\enquote {\bibinfo {title}
  {{Rediscovering orbital mechanics with machine learning}},}\ }\href@noop {}
  {\  (\bibinfo {year} {2022})},\ \Eprint {http://arxiv.org/abs/2202.02306}
  {arXiv:2202.02306 [astro-ph.EP]} \BibitemShut {NoStop}%
\bibitem [{\citenamefont {{Matchev}}\ \emph {et~al.}(2022)\citenamefont
  {{Matchev}}, \citenamefont {{Matcheva}},\ and\ \citenamefont
  {{Roman}}}]{Matchev2022ApJ}%
  \BibitemOpen
  \bibfield  {author} {\bibinfo {author} {\bibfnamefont {Konstantin~T.}\
  \bibnamefont {{Matchev}}}, \bibinfo {author} {\bibfnamefont {Katia}\
  \bibnamefont {{Matcheva}}}, \ and\ \bibinfo {author} {\bibfnamefont
  {Alexander}\ \bibnamefont {{Roman}}},\ }\bibfield  {title} {\enquote
  {\bibinfo {title} {{Analytical Modeling of Exoplanet Transit Spectroscopy
  with Dimensional Analysis and Symbolic Regression}},}\ }\href {\doibase
  10.3847/1538-4357/ac610c} {\bibfield  {journal} {\bibinfo  {journal} {\apj}\
  }\textbf {\bibinfo {volume} {930}},\ \bibinfo {eid} {33} (\bibinfo {year}
  {2022})},\ \Eprint {http://arxiv.org/abs/2112.11600} {arXiv:2112.11600
  [astro-ph.EP]} \BibitemShut {NoStop}%
\bibitem [{\citenamefont {Choi}(2011)}]{Choi:2010wa}%
  \BibitemOpen
  \bibfield  {author} {\bibinfo {author} {\bibfnamefont {Suyong}\ \bibnamefont
  {Choi}},\ }\bibfield  {title} {\enquote {\bibinfo {title} {{Construction of a
  Kinematic Variable Sensitive to the Mass of the Standard Model Higgs Boson in
  $H -> WW^* -> l^+ \nu l^- \bar{\nu}$ using Symbolic Regression}},}\ }\href
  {\doibase 10.1007/JHEP08(2011)110} {\bibfield  {journal} {\bibinfo  {journal}
  {JHEP}\ }\textbf {\bibinfo {volume} {08}},\ \bibinfo {pages} {110} (\bibinfo
  {year} {2011})},\ \Eprint {http://arxiv.org/abs/1006.4998} {arXiv:1006.4998
  [hep-ph]} \BibitemShut {NoStop}%
\bibitem [{\citenamefont {Butter}\ \emph {et~al.}(2021)\citenamefont {Butter},
  \citenamefont {Plehn}, \citenamefont {Soybelman},\ and\ \citenamefont
  {Brehmer}}]{Butter:2021rvz}%
  \BibitemOpen
  \bibfield  {author} {\bibinfo {author} {\bibfnamefont {Anja}\ \bibnamefont
  {Butter}}, \bibinfo {author} {\bibfnamefont {Tilman}\ \bibnamefont {Plehn}},
  \bibinfo {author} {\bibfnamefont {Nathalie}\ \bibnamefont {Soybelman}}, \
  and\ \bibinfo {author} {\bibfnamefont {Johann}\ \bibnamefont {Brehmer}},\
  }\bibfield  {title} {\enquote {\bibinfo {title} {{Back to the Formula -- LHC
  Edition}},}\ }\href@noop {} {\  (\bibinfo {year} {2021})},\ \Eprint
  {http://arxiv.org/abs/2109.10414} {arXiv:2109.10414 [hep-ph]} \BibitemShut
  {NoStop}%
\bibitem [{\citenamefont {Dersy}\ \emph {et~al.}(2022)\citenamefont {Dersy},
  \citenamefont {Schwartz},\ and\ \citenamefont {Zhang}}]{Dersy:2022bym}%
  \BibitemOpen
  \bibfield  {author} {\bibinfo {author} {\bibfnamefont {Aur\'elien}\
  \bibnamefont {Dersy}}, \bibinfo {author} {\bibfnamefont {Matthew~D.}\
  \bibnamefont {Schwartz}}, \ and\ \bibinfo {author} {\bibfnamefont {Xiaoyuan}\
  \bibnamefont {Zhang}},\ }\bibfield  {title} {\enquote {\bibinfo {title}
  {{Simplifying Polylogarithms with Machine Learning}},}\ }\href@noop {} {\
  (\bibinfo {year} {2022})},\ \Eprint {http://arxiv.org/abs/2206.04115}
  {arXiv:2206.04115 [cs.LG]} \BibitemShut {NoStop}%
\bibitem [{\citenamefont {Alnuqaydan}\ \emph {et~al.}(2022)\citenamefont
  {Alnuqaydan}, \citenamefont {Gleyzer},\ and\ \citenamefont
  {Prosper}}]{Alnuqaydan:2022ncd}%
  \BibitemOpen
  \bibfield  {author} {\bibinfo {author} {\bibfnamefont {Abdulhakim}\
  \bibnamefont {Alnuqaydan}}, \bibinfo {author} {\bibfnamefont {Sergei}\
  \bibnamefont {Gleyzer}}, \ and\ \bibinfo {author} {\bibfnamefont {Harrison}\
  \bibnamefont {Prosper}},\ }\bibfield  {title} {\enquote {\bibinfo {title}
  {{SYMBA: Symbolic Computation of Squared Amplitudes in High Energy Physics
  with Machine ALearning}},}\ }\href@noop {} {\  (\bibinfo {year} {2022})},\
  \Eprint {http://arxiv.org/abs/2206.08901} {arXiv:2206.08901 [hep-ph]}
  \BibitemShut {NoStop}%
\bibitem [{\citenamefont {Wang}\ \emph {et~al.}(2019)\citenamefont {Wang},
  \citenamefont {Wagner},\ and\ \citenamefont
  {Rondinelli}}]{wang_wagner_rondinelli_2019}%
  \BibitemOpen
  \bibfield  {author} {\bibinfo {author} {\bibfnamefont {Yiqun}\ \bibnamefont
  {Wang}}, \bibinfo {author} {\bibfnamefont {Nicholas}\ \bibnamefont {Wagner}},
  \ and\ \bibinfo {author} {\bibfnamefont {James~M.}\ \bibnamefont
  {Rondinelli}},\ }\bibfield  {title} {\enquote {\bibinfo {title} {Symbolic
  regression in materials science},}\ }\href {\doibase 10.1557/mrc.2019.85}
  {\bibfield  {journal} {\bibinfo  {journal} {MRS Communications}\ }\textbf
  {\bibinfo {volume} {9}},\ \bibinfo {pages} {793–805} (\bibinfo {year}
  {2019})}\BibitemShut {NoStop}%
\bibitem [{\citenamefont {Arechiga}\ \emph {et~al.}(2021)\citenamefont
  {Arechiga}, \citenamefont {Chen}, \citenamefont {Chen}, \citenamefont
  {Zhang}, \citenamefont {Iliev}, \citenamefont {Toyoda},\ and\ \citenamefont
  {Lyons}}]{Arechiga2021}%
  \BibitemOpen
  \bibfield  {author} {\bibinfo {author} {\bibfnamefont {Nikos}\ \bibnamefont
  {Arechiga}}, \bibinfo {author} {\bibfnamefont {Francine}\ \bibnamefont
  {Chen}}, \bibinfo {author} {\bibfnamefont {Yan-Ying}\ \bibnamefont {Chen}},
  \bibinfo {author} {\bibfnamefont {Yanxia}\ \bibnamefont {Zhang}}, \bibinfo
  {author} {\bibfnamefont {Rumen}\ \bibnamefont {Iliev}}, \bibinfo {author}
  {\bibfnamefont {Heishiro}\ \bibnamefont {Toyoda}}, \ and\ \bibinfo {author}
  {\bibfnamefont {Kent}\ \bibnamefont {Lyons}},\ }\href@noop {} {\enquote
  {\bibinfo {title} {Accelerating understanding of scientific experiments with
  end to end symbolic regression},}\ } (\bibinfo {year} {2021}),\ \Eprint
  {http://arxiv.org/abs/2112.04023} {arXiv:2112.04023 [cs.LG]} \BibitemShut
  {NoStop}%
\bibitem [{\citenamefont {Phillips}\ \emph {et~al.}(2021)\citenamefont
  {Phillips}, \citenamefont {Hahn}, \citenamefont {Fontana}, \citenamefont
  {Yates}, \citenamefont {Greene}, \citenamefont {Broniatowski},\ and\
  \citenamefont {Przybocki}}]{559461}%
  \BibitemOpen
  \bibfield  {author} {\bibinfo {author} {\bibfnamefont {P.}~\bibnamefont
  {Phillips}}, \bibinfo {author} {\bibfnamefont {Carina}\ \bibnamefont {Hahn}},
  \bibinfo {author} {\bibfnamefont {Peter}\ \bibnamefont {Fontana}}, \bibinfo
  {author} {\bibfnamefont {Amy}\ \bibnamefont {Yates}}, \bibinfo {author}
  {\bibfnamefont {Kristen}\ \bibnamefont {Greene}}, \bibinfo {author}
  {\bibfnamefont {David}\ \bibnamefont {Broniatowski}}, \ and\ \bibinfo
  {author} {\bibfnamefont {Mark}\ \bibnamefont {Przybocki}},\ }\href {\doibase
  https://doi.org/10.6028/NIST.IR.8312} {\enquote {\bibinfo {title} {Four
  principles of explainable artificial intelligence},}\ } (\bibinfo {year}
  {2021}),\ \Eprint {http://arxiv.org/abs/NIST Interagency/Internal Report
  (NISTIR), National Institute of Standards and Technology, Gaithersburg, MD}
  {NIST Interagency/Internal Report (NISTIR), National Institute of Standards
  and Technology, Gaithersburg, MD} \BibitemShut {NoStop}%
\bibitem [{\citenamefont {Fajardo-Fontiveros}\ \emph
  {et~al.}(2022)\citenamefont {Fajardo-Fontiveros}, \citenamefont {Reichardt},
  \citenamefont {Rios}, \citenamefont {Duch}, \citenamefont {Sales-Pardo},\
  and\ \citenamefont {Guimera}}]{2204.02704}%
  \BibitemOpen
  \bibfield  {author} {\bibinfo {author} {\bibfnamefont {Oscar}\ \bibnamefont
  {Fajardo-Fontiveros}}, \bibinfo {author} {\bibfnamefont {Ignasi}\
  \bibnamefont {Reichardt}}, \bibinfo {author} {\bibfnamefont {Harry R.
  De~Los}\ \bibnamefont {Rios}}, \bibinfo {author} {\bibfnamefont {Jordi}\
  \bibnamefont {Duch}}, \bibinfo {author} {\bibfnamefont {Marta}\ \bibnamefont
  {Sales-Pardo}}, \ and\ \bibinfo {author} {\bibfnamefont {Roger}\ \bibnamefont
  {Guimera}},\ }\href {\doibase 10.48550/ARXIV.2204.02704} {\enquote {\bibinfo
  {title} {Fundamental limits to learning closed-form mathematical models from
  data},}\ } (\bibinfo {year} {2022}),\ \Eprint
  {http://arxiv.org/abs/2204.02704} {arXiv:2204.02704 [cs.LG]} \BibitemShut
  {NoStop}%
\bibitem [{\citenamefont {Cranmer}(2020)}]{pysr}%
  \BibitemOpen
  \bibfield  {author} {\bibinfo {author} {\bibfnamefont {Miles}\ \bibnamefont
  {Cranmer}},\ }\href {\doibase 10.5281/zenodo.4041459} {\enquote {\bibinfo
  {title} {Pysr: Fast \& parallelized symbolic regression in python/julia},}\ }
  (\bibinfo {year} {2020}),\ \Eprint
  {http://arxiv.org/abs/http://doi.org/10.5281/zenodo.4041459}
  {http://doi.org/10.5281/zenodo.4041459} \BibitemShut {NoStop}%
\bibitem [{\citenamefont {Han}(2005)}]{Han:2005mu}%
  \BibitemOpen
  \bibfield  {author} {\bibinfo {author} {\bibfnamefont {Tao}\ \bibnamefont
  {Han}},\ }\bibfield  {title} {\enquote {\bibinfo {title} {{Collider
  phenomenology: Basic knowledge and techniques}},}\ }in\ \href {\doibase
  10.1142/9789812773579_0008} {\emph {\bibinfo {booktitle} {{Theoretical
  Advanced Study Institute in Elementary Particle Physics}: {Physics in D
  $\geqq$ 4}}}}\ (\bibinfo {year} {2005})\ pp.\ \bibinfo {pages} {407--454},\
  \Eprint {http://arxiv.org/abs/hep-ph/0508097} {arXiv:hep-ph/0508097}
  \BibitemShut {NoStop}%
\bibitem [{\citenamefont {Barr}\ and\ \citenamefont
  {Lester}(2010)}]{Barr:2010zj}%
  \BibitemOpen
  \bibfield  {author} {\bibinfo {author} {\bibfnamefont {Alan~J.}\ \bibnamefont
  {Barr}}\ and\ \bibinfo {author} {\bibfnamefont {Christopher~G.}\ \bibnamefont
  {Lester}},\ }\bibfield  {title} {\enquote {\bibinfo {title} {{A Review of the
  Mass Measurement Techniques proposed for the Large Hadron Collider}},}\
  }\href {\doibase 10.1088/0954-3899/37/12/123001} {\bibfield  {journal}
  {\bibinfo  {journal} {J. Phys. G}\ }\textbf {\bibinfo {volume} {37}},\
  \bibinfo {pages} {123001} (\bibinfo {year} {2010})},\ \Eprint
  {http://arxiv.org/abs/1004.2732} {arXiv:1004.2732 [hep-ph]} \BibitemShut
  {NoStop}%
\bibitem [{\citenamefont {Barr}\ \emph {et~al.}(2011)\citenamefont {Barr},
  \citenamefont {Khoo}, \citenamefont {Konar}, \citenamefont {Kong},
  \citenamefont {Lester}, \citenamefont {Matchev},\ and\ \citenamefont
  {Park}}]{Barr:2011xt}%
  \BibitemOpen
  \bibfield  {author} {\bibinfo {author} {\bibfnamefont {A.~J.}\ \bibnamefont
  {Barr}}, \bibinfo {author} {\bibfnamefont {T.~J.}\ \bibnamefont {Khoo}},
  \bibinfo {author} {\bibfnamefont {P.}~\bibnamefont {Konar}}, \bibinfo
  {author} {\bibfnamefont {K.}~\bibnamefont {Kong}}, \bibinfo {author}
  {\bibfnamefont {C.~G.}\ \bibnamefont {Lester}}, \bibinfo {author}
  {\bibfnamefont {K.~T.}\ \bibnamefont {Matchev}}, \ and\ \bibinfo {author}
  {\bibfnamefont {M.}~\bibnamefont {Park}},\ }\bibfield  {title} {\enquote
  {\bibinfo {title} {{Guide to transverse projections and mass-constraining
  variables}},}\ }\href {\doibase 10.1103/PhysRevD.84.095031} {\bibfield
  {journal} {\bibinfo  {journal} {Phys. Rev. D}\ }\textbf {\bibinfo {volume}
  {84}},\ \bibinfo {pages} {095031} (\bibinfo {year} {2011})},\ \Eprint
  {http://arxiv.org/abs/1105.2977} {arXiv:1105.2977 [hep-ph]} \BibitemShut
  {NoStop}%
\bibitem [{\citenamefont {Franceschini}\ \emph {et~al.}(2022)\citenamefont
  {Franceschini}, \citenamefont {Kim}, \citenamefont {Kong}, \citenamefont
  {Matchev}, \citenamefont {Park},\ and\ \citenamefont
  {Shyamsundar}}]{Franceschini:2022vck}%
  \BibitemOpen
  \bibfield  {author} {\bibinfo {author} {\bibfnamefont {Roberto}\ \bibnamefont
  {Franceschini}}, \bibinfo {author} {\bibfnamefont {Doojin}\ \bibnamefont
  {Kim}}, \bibinfo {author} {\bibfnamefont {Kyoungchul}\ \bibnamefont {Kong}},
  \bibinfo {author} {\bibfnamefont {Konstantin~T.}\ \bibnamefont {Matchev}},
  \bibinfo {author} {\bibfnamefont {Myeonghun}\ \bibnamefont {Park}}, \ and\
  \bibinfo {author} {\bibfnamefont {Prasanth}\ \bibnamefont {Shyamsundar}},\
  }\bibfield  {title} {\enquote {\bibinfo {title} {{Kinematic Variables and
  Feature Engineering for Particle Phenomenology}},}\ }\href@noop {} {\
  (\bibinfo {year} {2022})},\ \Eprint {http://arxiv.org/abs/2206.13431}
  {arXiv:2206.13431 [hep-ph]} \BibitemShut {NoStop}%
\bibitem [{\citenamefont {Banfi}\ \emph {et~al.}(2010)\citenamefont {Banfi},
  \citenamefont {Salam},\ and\ \citenamefont {Zanderighi}}]{Banfi:2010xy}%
  \BibitemOpen
  \bibfield  {author} {\bibinfo {author} {\bibfnamefont {Andrea}\ \bibnamefont
  {Banfi}}, \bibinfo {author} {\bibfnamefont {Gavin~P.}\ \bibnamefont {Salam}},
  \ and\ \bibinfo {author} {\bibfnamefont {Giulia}\ \bibnamefont
  {Zanderighi}},\ }\bibfield  {title} {\enquote {\bibinfo {title}
  {{Phenomenology of event shapes at hadron colliders}},}\ }\href {\doibase
  10.1007/JHEP06(2010)038} {\bibfield  {journal} {\bibinfo  {journal} {JHEP}\
  }\textbf {\bibinfo {volume} {06}},\ \bibinfo {pages} {038} (\bibinfo {year}
  {2010})},\ \Eprint {http://arxiv.org/abs/1001.4082} {arXiv:1001.4082
  [hep-ph]} \BibitemShut {NoStop}%
\bibitem [{\citenamefont {Stewart}\ \emph {et~al.}(2010)\citenamefont
  {Stewart}, \citenamefont {Tackmann},\ and\ \citenamefont
  {Waalewijn}}]{Stewart:2010tn}%
  \BibitemOpen
  \bibfield  {author} {\bibinfo {author} {\bibfnamefont {Iain~W.}\ \bibnamefont
  {Stewart}}, \bibinfo {author} {\bibfnamefont {Frank~J.}\ \bibnamefont
  {Tackmann}}, \ and\ \bibinfo {author} {\bibfnamefont {Wouter~J.}\
  \bibnamefont {Waalewijn}},\ }\bibfield  {title} {\enquote {\bibinfo {title}
  {{N-Jettiness: An Inclusive Event Shape to Veto Jets}},}\ }\href {\doibase
  10.1103/PhysRevLett.105.092002} {\bibfield  {journal} {\bibinfo  {journal}
  {Phys. Rev. Lett.}\ }\textbf {\bibinfo {volume} {105}},\ \bibinfo {pages}
  {092002} (\bibinfo {year} {2010})},\ \Eprint {http://arxiv.org/abs/1004.2489}
  {arXiv:1004.2489 [hep-ph]} \BibitemShut {NoStop}%
\bibitem [{\citenamefont {Thaler}\ and\ \citenamefont
  {Van~Tilburg}(2011)}]{Thaler:2010tr}%
  \BibitemOpen
  \bibfield  {author} {\bibinfo {author} {\bibfnamefont {Jesse}\ \bibnamefont
  {Thaler}}\ and\ \bibinfo {author} {\bibfnamefont {Ken}\ \bibnamefont
  {Van~Tilburg}},\ }\bibfield  {title} {\enquote {\bibinfo {title}
  {{Identifying Boosted Objects with N-subjettiness}},}\ }\href {\doibase
  10.1007/JHEP03(2011)015} {\bibfield  {journal} {\bibinfo  {journal} {JHEP}\
  }\textbf {\bibinfo {volume} {03}},\ \bibinfo {pages} {015} (\bibinfo {year}
  {2011})},\ \Eprint {http://arxiv.org/abs/1011.2268} {arXiv:1011.2268
  [hep-ph]} \BibitemShut {NoStop}%
\bibitem [{\citenamefont {Lester}\ and\ \citenamefont
  {Summers}(1999)}]{Lester:1999tx}%
  \BibitemOpen
  \bibfield  {author} {\bibinfo {author} {\bibfnamefont {C.~G.}\ \bibnamefont
  {Lester}}\ and\ \bibinfo {author} {\bibfnamefont {D.~J.}\ \bibnamefont
  {Summers}},\ }\bibfield  {title} {\enquote {\bibinfo {title} {{Measuring
  masses of semiinvisibly decaying particles pair produced at hadron
  colliders}},}\ }\href {\doibase 10.1016/S0370-2693(99)00945-4} {\bibfield
  {journal} {\bibinfo  {journal} {Phys. Lett. B}\ }\textbf {\bibinfo {volume}
  {463}},\ \bibinfo {pages} {99--103} (\bibinfo {year} {1999})},\ \Eprint
  {http://arxiv.org/abs/hep-ph/9906349} {arXiv:hep-ph/9906349} \BibitemShut
  {NoStop}%
\bibitem [{\citenamefont {Barr}\ \emph {et~al.}(2003)\citenamefont {Barr},
  \citenamefont {Lester},\ and\ \citenamefont {Stephens}}]{Barr:2003rg}%
  \BibitemOpen
  \bibfield  {author} {\bibinfo {author} {\bibfnamefont {Alan}\ \bibnamefont
  {Barr}}, \bibinfo {author} {\bibfnamefont {Christopher}\ \bibnamefont
  {Lester}}, \ and\ \bibinfo {author} {\bibfnamefont {P.}~\bibnamefont
  {Stephens}},\ }\bibfield  {title} {\enquote {\bibinfo {title} {{m(T2): The
  Truth behind the glamour}},}\ }\href {\doibase 10.1088/0954-3899/29/10/304}
  {\bibfield  {journal} {\bibinfo  {journal} {J. Phys. G}\ }\textbf {\bibinfo
  {volume} {29}},\ \bibinfo {pages} {2343--2363} (\bibinfo {year} {2003})},\
  \Eprint {http://arxiv.org/abs/hep-ph/0304226} {arXiv:hep-ph/0304226}
  \BibitemShut {NoStop}%
\bibitem [{\citenamefont {Cho}\ \emph {et~al.}(2014)\citenamefont {Cho},
  \citenamefont {Gainer}, \citenamefont {Kim}, \citenamefont {Matchev},
  \citenamefont {Moortgat}, \citenamefont {Pape},\ and\ \citenamefont
  {Park}}]{Cho:2014naa}%
  \BibitemOpen
  \bibfield  {author} {\bibinfo {author} {\bibfnamefont {Won~Sang}\
  \bibnamefont {Cho}}, \bibinfo {author} {\bibfnamefont {James~S.}\
  \bibnamefont {Gainer}}, \bibinfo {author} {\bibfnamefont {Doojin}\
  \bibnamefont {Kim}}, \bibinfo {author} {\bibfnamefont {Konstantin~T.}\
  \bibnamefont {Matchev}}, \bibinfo {author} {\bibfnamefont {Filip}\
  \bibnamefont {Moortgat}}, \bibinfo {author} {\bibfnamefont {Luc}\
  \bibnamefont {Pape}}, \ and\ \bibinfo {author} {\bibfnamefont {Myeonghun}\
  \bibnamefont {Park}},\ }\bibfield  {title} {\enquote {\bibinfo {title}
  {{On-shell constrained $M_2$ variables with applications to mass measurements
  and topology disambiguation}},}\ }\href {\doibase 10.1007/JHEP08(2014)070}
  {\bibfield  {journal} {\bibinfo  {journal} {JHEP}\ }\textbf {\bibinfo
  {volume} {08}},\ \bibinfo {pages} {070} (\bibinfo {year} {2014})},\ \Eprint
  {http://arxiv.org/abs/1401.1449} {arXiv:1401.1449 [hep-ph]} \BibitemShut
  {NoStop}%
\bibitem [{\citenamefont {Cho}\ \emph {et~al.}(2016)\citenamefont {Cho},
  \citenamefont {Gainer}, \citenamefont {Kim}, \citenamefont {Lim},
  \citenamefont {Matchev}, \citenamefont {Moortgat}, \citenamefont {Pape},\
  and\ \citenamefont {Park}}]{Cho:2015laa}%
  \BibitemOpen
  \bibfield  {author} {\bibinfo {author} {\bibfnamefont {Won~Sang}\
  \bibnamefont {Cho}}, \bibinfo {author} {\bibfnamefont {James~S.}\
  \bibnamefont {Gainer}}, \bibinfo {author} {\bibfnamefont {Doojin}\
  \bibnamefont {Kim}}, \bibinfo {author} {\bibfnamefont {Sung~Hak}\
  \bibnamefont {Lim}}, \bibinfo {author} {\bibfnamefont {Konstantin~T.}\
  \bibnamefont {Matchev}}, \bibinfo {author} {\bibfnamefont {Filip}\
  \bibnamefont {Moortgat}}, \bibinfo {author} {\bibfnamefont {Luc}\
  \bibnamefont {Pape}}, \ and\ \bibinfo {author} {\bibfnamefont {Myeonghun}\
  \bibnamefont {Park}},\ }\bibfield  {title} {\enquote {\bibinfo {title}
  {{OPTIMASS: A Package for the Minimization of Kinematic Mass Functions with
  Constraints}},}\ }\href {\doibase 10.1007/JHEP01(2016)026} {\bibfield
  {journal} {\bibinfo  {journal} {JHEP}\ }\textbf {\bibinfo {volume} {01}},\
  \bibinfo {pages} {026} (\bibinfo {year} {2016})},\ \Eprint
  {http://arxiv.org/abs/1508.00589} {arXiv:1508.00589 [hep-ph]} \BibitemShut
  {NoStop}%
\bibitem [{\citenamefont {Cho}\ \emph {et~al.}(2015)\citenamefont {Cho},
  \citenamefont {Gainer}, \citenamefont {Kim}, \citenamefont {Matchev},
  \citenamefont {Moortgat}, \citenamefont {Pape},\ and\ \citenamefont
  {Park}}]{Cho:2014yma}%
  \BibitemOpen
  \bibfield  {author} {\bibinfo {author} {\bibfnamefont {Won~Sang}\
  \bibnamefont {Cho}}, \bibinfo {author} {\bibfnamefont {James~S.}\
  \bibnamefont {Gainer}}, \bibinfo {author} {\bibfnamefont {Doojin}\
  \bibnamefont {Kim}}, \bibinfo {author} {\bibfnamefont {Konstantin~T.}\
  \bibnamefont {Matchev}}, \bibinfo {author} {\bibfnamefont {Filip}\
  \bibnamefont {Moortgat}}, \bibinfo {author} {\bibfnamefont {Luc}\
  \bibnamefont {Pape}}, \ and\ \bibinfo {author} {\bibfnamefont {Myeonghun}\
  \bibnamefont {Park}},\ }\bibfield  {title} {\enquote {\bibinfo {title}
  {{Improving the sensitivity of stop searches with on-shell constrained
  invariant mass variables}},}\ }\href {\doibase 10.1007/JHEP05(2015)040}
  {\bibfield  {journal} {\bibinfo  {journal} {JHEP}\ }\textbf {\bibinfo
  {volume} {05}},\ \bibinfo {pages} {040} (\bibinfo {year} {2015})},\ \Eprint
  {http://arxiv.org/abs/1411.0664} {arXiv:1411.0664 [hep-ph]} \BibitemShut
  {NoStop}%
\bibitem [{\citenamefont {Ross}\ and\ \citenamefont
  {Serna}(2008)}]{Ross:2007rm}%
  \BibitemOpen
  \bibfield  {author} {\bibinfo {author} {\bibfnamefont {Graham~G.}\
  \bibnamefont {Ross}}\ and\ \bibinfo {author} {\bibfnamefont {Mario}\
  \bibnamefont {Serna}},\ }\bibfield  {title} {\enquote {\bibinfo {title}
  {{Mass determination of new states at hadron colliders}},}\ }\href {\doibase
  10.1016/j.physletb.2008.06.003} {\bibfield  {journal} {\bibinfo  {journal}
  {Phys. Lett. B}\ }\textbf {\bibinfo {volume} {665}},\ \bibinfo {pages}
  {212--218} (\bibinfo {year} {2008})},\ \Eprint
  {http://arxiv.org/abs/0712.0943} {arXiv:0712.0943 [hep-ph]} \BibitemShut
  {NoStop}%
\bibitem [{\citenamefont {Lester}(2011)}]{Lester:2011nj}%
  \BibitemOpen
  \bibfield  {author} {\bibinfo {author} {\bibfnamefont {Christopher~G.}\
  \bibnamefont {Lester}},\ }\bibfield  {title} {\enquote {\bibinfo {title}
  {{The stransverse mass, MT2, in special cases}},}\ }\href {\doibase
  10.1007/JHEP05(2011)076} {\bibfield  {journal} {\bibinfo  {journal} {JHEP}\
  }\textbf {\bibinfo {volume} {05}},\ \bibinfo {pages} {076} (\bibinfo {year}
  {2011})},\ \Eprint {http://arxiv.org/abs/1103.5682} {arXiv:1103.5682
  [hep-ph]} \BibitemShut {NoStop}%
\bibitem [{\citenamefont {Lally}\ and\ \citenamefont
  {Lester}(2012)}]{Lally:2012uj}%
  \BibitemOpen
  \bibfield  {author} {\bibinfo {author} {\bibfnamefont {Colin~H.}\
  \bibnamefont {Lally}}\ and\ \bibinfo {author} {\bibfnamefont
  {Christopher~G.}\ \bibnamefont {Lester}},\ }\bibfield  {title} {\enquote
  {\bibinfo {title} {{Properties of MT2 in the massless limit}},}\ }\href@noop
  {} {\  (\bibinfo {year} {2012})},\ \Eprint {http://arxiv.org/abs/1211.1542}
  {arXiv:1211.1542 [hep-ph]} \BibitemShut {NoStop}%
\bibitem [{\citenamefont {Lester}\ and\ \citenamefont
  {Barr}(2007)}]{Lester:2007fq}%
  \BibitemOpen
  \bibfield  {author} {\bibinfo {author} {\bibfnamefont {Christopher}\
  \bibnamefont {Lester}}\ and\ \bibinfo {author} {\bibfnamefont {Alan}\
  \bibnamefont {Barr}},\ }\bibfield  {title} {\enquote {\bibinfo {title}
  {{MTGEN: Mass scale measurements in pair-production at colliders}},}\ }\href
  {\doibase 10.1088/1126-6708/2007/12/102} {\bibfield  {journal} {\bibinfo
  {journal} {JHEP}\ }\textbf {\bibinfo {volume} {12}},\ \bibinfo {pages} {102}
  (\bibinfo {year} {2007})},\ \Eprint {http://arxiv.org/abs/0708.1028}
  {arXiv:0708.1028 [hep-ph]} \BibitemShut {NoStop}%
\bibitem [{\citenamefont {Cho}\ \emph {et~al.}(2008{\natexlab{a}})\citenamefont
  {Cho}, \citenamefont {Choi}, \citenamefont {Kim},\ and\ \citenamefont
  {Park}}]{Cho:2007qv}%
  \BibitemOpen
  \bibfield  {author} {\bibinfo {author} {\bibfnamefont {Won~Sang}\
  \bibnamefont {Cho}}, \bibinfo {author} {\bibfnamefont {Kiwoon}\ \bibnamefont
  {Choi}}, \bibinfo {author} {\bibfnamefont {Yeong~Gyun}\ \bibnamefont {Kim}},
  \ and\ \bibinfo {author} {\bibfnamefont {Chan~Beom}\ \bibnamefont {Park}},\
  }\bibfield  {title} {\enquote {\bibinfo {title} {{Gluino Stransverse
  Mass}},}\ }\href {\doibase 10.1103/PhysRevLett.100.171801} {\bibfield
  {journal} {\bibinfo  {journal} {Phys. Rev. Lett.}\ }\textbf {\bibinfo
  {volume} {100}},\ \bibinfo {pages} {171801} (\bibinfo {year}
  {2008}{\natexlab{a}})},\ \Eprint {http://arxiv.org/abs/0709.0288}
  {arXiv:0709.0288 [hep-ph]} \BibitemShut {NoStop}%
\bibitem [{\citenamefont {Cho}\ \emph {et~al.}(2008{\natexlab{b}})\citenamefont
  {Cho}, \citenamefont {Choi}, \citenamefont {Kim},\ and\ \citenamefont
  {Park}}]{Cho:2007dh}%
  \BibitemOpen
  \bibfield  {author} {\bibinfo {author} {\bibfnamefont {Won~Sang}\
  \bibnamefont {Cho}}, \bibinfo {author} {\bibfnamefont {Kiwoon}\ \bibnamefont
  {Choi}}, \bibinfo {author} {\bibfnamefont {Yeong~Gyun}\ \bibnamefont {Kim}},
  \ and\ \bibinfo {author} {\bibfnamefont {Chan~Beom}\ \bibnamefont {Park}},\
  }\bibfield  {title} {\enquote {\bibinfo {title} {{Measuring superparticle
  masses at hadron collider using the transverse mass kink}},}\ }\href
  {\doibase 10.1088/1126-6708/2008/02/035} {\bibfield  {journal} {\bibinfo
  {journal} {JHEP}\ }\textbf {\bibinfo {volume} {02}},\ \bibinfo {pages} {035}
  (\bibinfo {year} {2008}{\natexlab{b}})},\ \Eprint
  {http://arxiv.org/abs/0711.4526} {arXiv:0711.4526 [hep-ph]} \BibitemShut
  {NoStop}%
\bibitem [{\citenamefont {Lester}\ and\ \citenamefont
  {Nachman}(2015)}]{Lester:2014yga}%
  \BibitemOpen
  \bibfield  {author} {\bibinfo {author} {\bibfnamefont {Christopher~G.}\
  \bibnamefont {Lester}}\ and\ \bibinfo {author} {\bibfnamefont {Benjamin}\
  \bibnamefont {Nachman}},\ }\bibfield  {title} {\enquote {\bibinfo {title}
  {{Bisection-based asymmetric M$_{T2}$ computation: a higher precision
  calculator than existing symmetric methods}},}\ }\href {\doibase
  10.1007/JHEP03(2015)100} {\bibfield  {journal} {\bibinfo  {journal} {JHEP}\
  }\textbf {\bibinfo {volume} {03}},\ \bibinfo {pages} {100} (\bibinfo {year}
  {2015})},\ \Eprint {http://arxiv.org/abs/1411.4312} {arXiv:1411.4312
  [hep-ph]} \BibitemShut {NoStop}%
\bibitem [{\citenamefont {Birkedal}\ \emph {et~al.}(2004)\citenamefont
  {Birkedal}, \citenamefont {Matchev},\ and\ \citenamefont
  {Perelstein}}]{Birkedal:2004xn}%
  \BibitemOpen
  \bibfield  {author} {\bibinfo {author} {\bibfnamefont {Andreas}\ \bibnamefont
  {Birkedal}}, \bibinfo {author} {\bibfnamefont {Konstantin}\ \bibnamefont
  {Matchev}}, \ and\ \bibinfo {author} {\bibfnamefont {Maxim}\ \bibnamefont
  {Perelstein}},\ }\bibfield  {title} {\enquote {\bibinfo {title} {{Dark matter
  at colliders: A Model independent approach}},}\ }\href {\doibase
  10.1103/PhysRevD.70.077701} {\bibfield  {journal} {\bibinfo  {journal} {Phys.
  Rev. D}\ }\textbf {\bibinfo {volume} {70}},\ \bibinfo {pages} {077701}
  (\bibinfo {year} {2004})},\ \Eprint {http://arxiv.org/abs/hep-ph/0403004}
  {arXiv:hep-ph/0403004} \BibitemShut {NoStop}%
\bibitem [{\citenamefont {Gopalakrishna}\ \emph {et~al.}(2001)\citenamefont
  {Gopalakrishna}, \citenamefont {Perelstein},\ and\ \citenamefont
  {Wells}}]{Gopalakrishna:2001iv}%
  \BibitemOpen
  \bibfield  {author} {\bibinfo {author} {\bibfnamefont {Shrihari}\
  \bibnamefont {Gopalakrishna}}, \bibinfo {author} {\bibfnamefont {Maxim}\
  \bibnamefont {Perelstein}}, \ and\ \bibinfo {author} {\bibfnamefont
  {James~D.}\ \bibnamefont {Wells}},\ }\bibfield  {title} {\enquote {\bibinfo
  {title} {{Extra dimensions versus supersymmetric interpretation of missing
  energy events at a linear collider}},}\ }\href@noop {} {\bibfield  {journal}
  {\bibinfo  {journal} {eConf}\ }\textbf {\bibinfo {volume} {C010630}},\
  \bibinfo {pages} {P311} (\bibinfo {year} {2001})},\ \Eprint
  {http://arxiv.org/abs/hep-ph/0110339} {arXiv:hep-ph/0110339} \BibitemShut
  {NoStop}%
\bibitem [{\citenamefont {Oller}\ \emph {et~al.}(2004)\citenamefont {Oller},
  \citenamefont {Eberl},\ and\ \citenamefont {Majerotto}}]{Oller:2004br}%
  \BibitemOpen
  \bibfield  {author} {\bibinfo {author} {\bibfnamefont {W.}~\bibnamefont
  {Oller}}, \bibinfo {author} {\bibfnamefont {H.}~\bibnamefont {Eberl}}, \ and\
  \bibinfo {author} {\bibfnamefont {W.}~\bibnamefont {Majerotto}},\ }\bibfield
  {title} {\enquote {\bibinfo {title} {{Full one loop corrections to neutralino
  pair production in e+ e- annihilation}},}\ }\href {\doibase
  10.1016/j.physletb.2004.03.075} {\bibfield  {journal} {\bibinfo  {journal}
  {Phys. Lett. B}\ }\textbf {\bibinfo {volume} {590}},\ \bibinfo {pages}
  {273--283} (\bibinfo {year} {2004})},\ \Eprint
  {http://arxiv.org/abs/hep-ph/0402134} {arXiv:hep-ph/0402134} \BibitemShut
  {NoStop}%
\bibitem [{\citenamefont {Mawatari}\ and\ \citenamefont
  {Oexl}(2014)}]{Mawatari:2014cja}%
  \BibitemOpen
  \bibfield  {author} {\bibinfo {author} {\bibfnamefont {Kentarou}\
  \bibnamefont {Mawatari}}\ and\ \bibinfo {author} {\bibfnamefont {Bettina}\
  \bibnamefont {Oexl}},\ }\bibfield  {title} {\enquote {\bibinfo {title}
  {{Monophoton signals in light gravitino production at $e^+ e^-$
  colliders}},}\ }\href {\doibase 10.1140/epjc/s10052-014-2909-0} {\bibfield
  {journal} {\bibinfo  {journal} {Eur. Phys. J. C}\ }\textbf {\bibinfo {volume}
  {74}},\ \bibinfo {pages} {2909} (\bibinfo {year} {2014})},\ \Eprint
  {http://arxiv.org/abs/1402.3223} {arXiv:1402.3223 [hep-ph]} \BibitemShut
  {NoStop}%
\bibitem [{\citenamefont {Kalinowski}\ \emph {et~al.}(2022)\citenamefont
  {Kalinowski}, \citenamefont {Mekala}, \citenamefont {Sopicki}, \citenamefont
  {Zarnecki},\ and\ \citenamefont {Kotlarski}}]{Kalinowski:2022xjw}%
  \BibitemOpen
  \bibfield  {author} {\bibinfo {author} {\bibfnamefont {J.}~\bibnamefont
  {Kalinowski}}, \bibinfo {author} {\bibfnamefont {K.}~\bibnamefont {Mekala}},
  \bibinfo {author} {\bibfnamefont {P.}~\bibnamefont {Sopicki}}, \bibinfo
  {author} {\bibfnamefont {A.~F.}\ \bibnamefont {Zarnecki}}, \ and\ \bibinfo
  {author} {\bibfnamefont {W.}~\bibnamefont {Kotlarski}},\ }\bibfield  {title}
  {\enquote {\bibinfo {title} {{Sensitivity of Future $e^+e^-$ Colliders to
  Processes of Dark Matter Production with Light Mediator Exchange}},}\ }\href
  {\doibase 10.5506/APhysPolBSupp.15.2-A10} {\bibfield  {journal} {\bibinfo
  {journal} {Acta Phys. Polon. Supp.}\ }\textbf {\bibinfo {volume} {15}},\
  \bibinfo {pages} {A10} (\bibinfo {year} {2022})}\BibitemShut {NoStop}%
\bibitem [{\citenamefont {Alwall}\ \emph {et~al.}(2011)\citenamefont {Alwall},
  \citenamefont {Herquet}, \citenamefont {Maltoni}, \citenamefont {Mattelaer},\
  and\ \citenamefont {Stelzer}}]{Alwall:2011uj}%
  \BibitemOpen
  \bibfield  {author} {\bibinfo {author} {\bibfnamefont {Johan}\ \bibnamefont
  {Alwall}}, \bibinfo {author} {\bibfnamefont {Michel}\ \bibnamefont
  {Herquet}}, \bibinfo {author} {\bibfnamefont {Fabio}\ \bibnamefont
  {Maltoni}}, \bibinfo {author} {\bibfnamefont {Olivier}\ \bibnamefont
  {Mattelaer}}, \ and\ \bibinfo {author} {\bibfnamefont {Tim}\ \bibnamefont
  {Stelzer}},\ }\bibfield  {title} {\enquote {\bibinfo {title} {{MadGraph 5 :
  Going Beyond}},}\ }\href {\doibase 10.1007/JHEP06(2011)128} {\bibfield
  {journal} {\bibinfo  {journal} {JHEP}\ }\textbf {\bibinfo {volume} {06}},\
  \bibinfo {pages} {128} (\bibinfo {year} {2011})},\ \Eprint
  {http://arxiv.org/abs/1106.0522} {arXiv:1106.0522 [hep-ph]} \BibitemShut
  {NoStop}%
\bibitem [{\citenamefont {Petersen}(2019)}]{DBLP:journals/corr/abs-1912-04871}%
  \BibitemOpen
  \bibfield  {author} {\bibinfo {author} {\bibfnamefont {Brenden~K.}\
  \bibnamefont {Petersen}},\ }\bibfield  {title} {\enquote {\bibinfo {title}
  {Deep symbolic regression: Recovering mathematical expressions from data via
  policy gradients},}\ }\href {http://arxiv.org/abs/1912.04871} {\bibfield
  {journal} {\bibinfo  {journal} {CoRR}\ }\textbf {\bibinfo {volume}
  {abs/1912.04871}} (\bibinfo {year} {2019})},\ \Eprint
  {http://arxiv.org/abs/1912.04871} {1912.04871} \BibitemShut {NoStop}%
\bibitem [{\citenamefont {Adloff}\ \emph {et~al.}(2009)\citenamefont {Adloff}
  \emph {et~al.}}]{CALICE:2008kht}%
  \BibitemOpen
  \bibfield  {author} {\bibinfo {author} {\bibfnamefont {C.}~\bibnamefont
  {Adloff}} \emph {et~al.} (\bibinfo {collaboration} {CALICE}),\ }\bibfield
  {title} {\enquote {\bibinfo {title} {{Response of the CALICE Si-W
  electromagnetic calorimeter physics prototype to electrons}},}\ }\href
  {\doibase 10.1016/j.nima.2009.07.026} {\bibfield  {journal} {\bibinfo
  {journal} {Nucl. Instrum. Meth. A}\ }\textbf {\bibinfo {volume} {608}},\
  \bibinfo {pages} {372--383} (\bibinfo {year} {2009})},\ \Eprint
  {http://arxiv.org/abs/0811.2354} {arXiv:0811.2354 [physics.ins-det]}
  \BibitemShut {NoStop}%
\bibitem [{\citenamefont {Bambade}\ \emph {et~al.}(2019)\citenamefont {Bambade}
  \emph {et~al.}}]{Bambade:2019fyw}%
  \BibitemOpen
  \bibfield  {author} {\bibinfo {author} {\bibfnamefont {Philip}\ \bibnamefont
  {Bambade}} \emph {et~al.},\ }\bibfield  {title} {\enquote {\bibinfo {title}
  {{The International Linear Collider: A Global Project}},}\ }\href@noop {} {\
  (\bibinfo {year} {2019})},\ \Eprint {http://arxiv.org/abs/1903.01629}
  {arXiv:1903.01629 [hep-ex]} \BibitemShut {NoStop}%
\bibitem [{\citenamefont {Habermehl}\ \emph {et~al.}(2020)\citenamefont
  {Habermehl}, \citenamefont {Berggren},\ and\ \citenamefont
  {List}}]{Habermehl:2020njb}%
  \BibitemOpen
  \bibfield  {author} {\bibinfo {author} {\bibfnamefont {Moritz}\ \bibnamefont
  {Habermehl}}, \bibinfo {author} {\bibfnamefont {Mikael}\ \bibnamefont
  {Berggren}}, \ and\ \bibinfo {author} {\bibfnamefont {Jenny}\ \bibnamefont
  {List}},\ }\bibfield  {title} {\enquote {\bibinfo {title} {{WIMP Dark Matter
  at the International Linear Collider}},}\ }\href {\doibase
  10.1103/PhysRevD.101.075053} {\bibfield  {journal} {\bibinfo  {journal}
  {Phys. Rev. D}\ }\textbf {\bibinfo {volume} {101}},\ \bibinfo {pages}
  {075053} (\bibinfo {year} {2020})},\ \Eprint
  {http://arxiv.org/abs/2001.03011} {arXiv:2001.03011 [hep-ex]} \BibitemShut
  {NoStop}%
\bibitem [{\citenamefont {Abramowicz}\ \emph {et~al.}(2020)\citenamefont
  {Abramowicz} \emph {et~al.}}]{ILDConceptGroup:2020sfq}%
  \BibitemOpen
  \bibfield  {author} {\bibinfo {author} {\bibfnamefont {Halina}\ \bibnamefont
  {Abramowicz}} \emph {et~al.} (\bibinfo {collaboration} {ILD Concept Group}),\
  }\bibfield  {title} {\enquote {\bibinfo {title} {{International Large
  Detector: Interim Design Report}},}\ }\href@noop {} {\  (\bibinfo {year}
  {2020})},\ \Eprint {http://arxiv.org/abs/2003.01116} {arXiv:2003.01116
  [physics.ins-det]} \BibitemShut {NoStop}%
\bibitem [{\citenamefont {Badger}\ \emph {et~al.}(2022)\citenamefont {Badger}
  \emph {et~al.}}]{Butter:2022rso}%
  \BibitemOpen
  \bibfield  {author} {\bibinfo {author} {\bibfnamefont {Simon}\ \bibnamefont
  {Badger}} \emph {et~al.},\ }\bibfield  {title} {\enquote {\bibinfo {title}
  {{Machine Learning and LHC Event Generation}},}\ }\href@noop {} {\  (\bibinfo
  {year} {2022})},\ \Eprint {http://arxiv.org/abs/2203.07460} {arXiv:2203.07460
  [hep-ph]} \BibitemShut {NoStop}%
\bibitem [{\citenamefont {Kawagoe}\ \emph {et~al.}(2005)\citenamefont
  {Kawagoe}, \citenamefont {Nojiri},\ and\ \citenamefont
  {Polesello}}]{Kawagoe:2004rz}%
  \BibitemOpen
  \bibfield  {author} {\bibinfo {author} {\bibfnamefont {K.}~\bibnamefont
  {Kawagoe}}, \bibinfo {author} {\bibfnamefont {M.~M.}\ \bibnamefont {Nojiri}},
  \ and\ \bibinfo {author} {\bibfnamefont {G.}~\bibnamefont {Polesello}},\
  }\bibfield  {title} {\enquote {\bibinfo {title} {{A New SUSY mass
  reconstruction method at the CERN LHC}},}\ }\href {\doibase
  10.1103/PhysRevD.71.035008} {\bibfield  {journal} {\bibinfo  {journal} {Phys.
  Rev. D}\ }\textbf {\bibinfo {volume} {71}},\ \bibinfo {pages} {035008}
  (\bibinfo {year} {2005})},\ \Eprint {http://arxiv.org/abs/hep-ph/0410160}
  {arXiv:hep-ph/0410160} \BibitemShut {NoStop}%
\bibitem [{\citenamefont {Burns}\ \emph {et~al.}(2009)\citenamefont {Burns},
  \citenamefont {Kong}, \citenamefont {Matchev},\ and\ \citenamefont
  {Park}}]{Burns:2008va}%
  \BibitemOpen
  \bibfield  {author} {\bibinfo {author} {\bibfnamefont {Michael}\ \bibnamefont
  {Burns}}, \bibinfo {author} {\bibfnamefont {Kyoungchul}\ \bibnamefont
  {Kong}}, \bibinfo {author} {\bibfnamefont {Konstantin~T.}\ \bibnamefont
  {Matchev}}, \ and\ \bibinfo {author} {\bibfnamefont {Myeonghun}\ \bibnamefont
  {Park}},\ }\bibfield  {title} {\enquote {\bibinfo {title} {{Using Subsystem
  MT2 for Complete Mass Determinations in Decay Chains with Missing Energy at
  Hadron Colliders}},}\ }\href {\doibase 10.1088/1126-6708/2009/03/143}
  {\bibfield  {journal} {\bibinfo  {journal} {JHEP}\ }\textbf {\bibinfo
  {volume} {03}},\ \bibinfo {pages} {143} (\bibinfo {year} {2009})},\ \Eprint
  {http://arxiv.org/abs/0810.5576} {arXiv:0810.5576 [hep-ph]} \BibitemShut
  {NoStop}%
\bibitem [{\citenamefont {d'Ascoli}\ \emph {et~al.}(2022)\citenamefont
  {d'Ascoli}, \citenamefont {Kamienny}, \citenamefont {Lample},\ and\
  \citenamefont {Charton}}]{https://doi.org/10.48550/arxiv.2201.04600}%
  \BibitemOpen
  \bibfield  {author} {\bibinfo {author} {\bibfnamefont {Stéphane}\
  \bibnamefont {d'Ascoli}}, \bibinfo {author} {\bibfnamefont
  {Pierre-Alexandre}\ \bibnamefont {Kamienny}}, \bibinfo {author}
  {\bibfnamefont {Guillaume}\ \bibnamefont {Lample}}, \ and\ \bibinfo {author}
  {\bibfnamefont {François}\ \bibnamefont {Charton}},\ }\href {\doibase
  10.48550/ARXIV.2201.04600} {\enquote {\bibinfo {title} {Deep symbolic
  regression for recurrent sequences},}\ } (\bibinfo {year} {2022}),\ \Eprint
  {http://arxiv.org/abs/2201.04600} {arXiv:2201.04600 [cs.LG]} \BibitemShut
  {NoStop}%
\bibitem [{\citenamefont {Kamienny}\ \emph {et~al.}(2022)\citenamefont
  {Kamienny}, \citenamefont {d'Ascoli}, \citenamefont {Lample},\ and\
  \citenamefont {Charton}}]{https://doi.org/10.48550/arxiv.2204.10532}%
  \BibitemOpen
  \bibfield  {author} {\bibinfo {author} {\bibfnamefont {Pierre-Alexandre}\
  \bibnamefont {Kamienny}}, \bibinfo {author} {\bibfnamefont {Stéphane}\
  \bibnamefont {d'Ascoli}}, \bibinfo {author} {\bibfnamefont {Guillaume}\
  \bibnamefont {Lample}}, \ and\ \bibinfo {author} {\bibfnamefont {François}\
  \bibnamefont {Charton}},\ }\href {\doibase 10.48550/ARXIV.2204.10532}
  {\enquote {\bibinfo {title} {End-to-end symbolic regression with
  transformers},}\ } (\bibinfo {year} {2022}),\ \Eprint
  {http://arxiv.org/abs/2204.10532} {arXiv:2204.10532 [cs.LG]} \BibitemShut
  {NoStop}%
\bibitem [{\citenamefont {Li}\ \emph {et~al.}(2022)\citenamefont {Li},
  \citenamefont {Yuan},\ and\ \citenamefont
  {Shen}}]{https://doi.org/10.48550/arxiv.2205.11798}%
  \BibitemOpen
  \bibfield  {author} {\bibinfo {author} {\bibfnamefont {Jiachen}\ \bibnamefont
  {Li}}, \bibinfo {author} {\bibfnamefont {Ye}~\bibnamefont {Yuan}}, \ and\
  \bibinfo {author} {\bibfnamefont {Hong-Bin}\ \bibnamefont {Shen}},\ }\href
  {\doibase 10.48550/ARXIV.2205.11798} {\enquote {\bibinfo {title} {Symbolic
  expression transformer: A computer vision approach for symbolic
  regression},}\ } (\bibinfo {year} {2022}),\ \Eprint
  {http://arxiv.org/abs/2205.11798} {arXiv:2205.11798 [cs.CV]} \BibitemShut
  {NoStop}%
\end{thebibliography}%

\end{document}